\documentclass[journal]{IEEEtran}


\usepackage{bigstrut}   
\usepackage[table]{xcolor} 
\usepackage{booktabs}
\usepackage{tabularx}        
\usepackage{makecell}        
\usepackage{array}         
\usepackage{graphicx}      
\usepackage[hidelinks]{hyperref}    
\usepackage{amsmath}
\usepackage{cleveref}      
\usepackage{cite}
\usepackage{subcaption} 
\usepackage[most]{tcolorbox}
\usepackage{listings}

\AtBeginDocument{%
  \providecommand\BibTeX{{%
    \normalfont B\kern-0.5em{\scshape i\kern-0.25em b}\kern-.8em\TeX}}}
\def\BibTeX{{\rm B\kern-.05em{\sc i\kern-.025em b}\kern-.08em
    T\kern-.1667em\lower.7ex\hbox{E}\kern-.125emX}}

\usepackage[nolist]{acronym}  

\usepackage{amsthm}
\newtheorem{example}{Example}
\newtheorem{resq}{Research Question}

\crefname{lstlisting}{listing}{listings}

\usepackage{etoolbox}

\definecolor{quotegreen}{rgb}{0.0,0.45,0.15}
\AtBeginEnvironment{quote}{\color{quotegreen}}

\newcounter{reviewer}
\newcounter{point}[reviewer]

\usepackage{titlesec}

\titlespacing*{\section}{0pt}{8pt plus 2pt minus 2pt}{4pt plus 1pt}
\titlespacing*{\subsection}{0pt}{6pt plus 2pt minus 2pt}{3pt plus 1pt}
\titlespacing*{\subsubsection}{0pt}{5pt plus 1pt minus 1pt}{2pt plus 1pt}

\usepackage[skip=4pt, font=footnotesize]{caption}
\newcommand{\revised}[1]{{#1}}  

\begin{document}
\begin{acronym}
\acro{IT}{information technology}
\acro{NetOps}{network operations}
\acro{IBN}{intent-based networking}
\acro{VLLM}{Vision-Language Large Model}
\acro{NLP}{natural language processing}
\acro{LLM}{large language model}
\acro{SMT}{satisfiability modulo theories}
\acro{AI}{artificial intelligence}
\acro{RNN}{recurrent neural network} 
\acro{DL}{deep learning}
\acro{CV}{computer vision}
\acro{VQA}{visual question answering}
\acro{SDN}{software-defined networking}
\acro{GNN}{graph neural network}
\acro{RAG}{retrieval augmented generation}
\acro{IG}{information gain}
\acro{TIUS}{topology image understanding score}
\acro{FSS}{fast screening score}
\acro{EVSS}{emulator validated screening score}
\acro{S3}{solution screening score}
\acro{S3-Z}{standardized solution screening score}
\acro{IIS}{Intent Implementation Score}
\acro{ReGeNet}{\textit{Reflective-GeNet}}
\end{acronym}

\title{Config But Verify! ReGeNet: A Reflective Multi-Agent Framework for Multimodal Intent-Based Networking}

\author{\begin{tabular}{c}
\normalsize Aviram Zilberman\textsuperscript{3,1,*}, Rubin Krief\textsuperscript{1}, Abed Showgan\textsuperscript{1}, Andres Murillo\textsuperscript{2}, Ortal Lavi\textsuperscript{2},\\
\normalsize Motoyoshi Sekiya\textsuperscript{2}, Beni Ifland\textsuperscript{1}, Asaf Shabtai\textsuperscript{1}, Yuval Elovici\textsuperscript{1}, Rami Puzis\textsuperscript{1}\\[4pt]
\small \textsuperscript{1}Ben-Gurion University of the Negev, Faculty of Computer and Information Science and Cyber@BGU\\
\small \textsuperscript{2}Fujitsu\\
\small \textsuperscript{3}Jerusalem College of Technology\\[3pt]
\small * Corresponding author: Aviram Zilberman (\href{mailto:aviramz@jct.ac.il}{aviramz@jct.ac.il})
\end{tabular}}

\maketitle
 
\begin{abstract}
Communication network engineering in enterprise environments is complex and error-prone, since configuration mistakes or overlooked topology changes can disrupt operations and introduce security risks. Automation research has mainly targeted textual, device-level configuration, giving less attention to topology updates. Recent \ac{IBN} approaches use \acp{LLM} to simplify network engineering, and best practice recommends automated verification to reduce erroneous or hallucinated outputs; existing verification mainly analyzes device configurations, overlooking topology-level intents needed for holistic frameworks. We introduce \textit{Reflective-GeNet (ReGeNet)}, a multi-agent network copilot with a self-verification loop that reduces \ac{LLM}-induced errors, extending \textit{GeNet}, a multimodal copilot interpreting device configurations and visual diagrams to modify networks per user intent. Experiments with nine \acp{LLM} on topology image interpretation show \textit{GPT-4.1-mini} and \textit{o4-mini} outperform vision-oriented \acp{LLM}; verification mainly benefits non-reasoning models, letting an efficient one match a costlier reasoning model, while reasoning models gain only marginally. Evaluations on a Cisco-inspired multimodal \ac{IBN} benchmark from GNS3Vault certification-style exercises show \textit{ReGeNet} improves reliability and trustworthiness, though these scenarios do not fully capture the scale, heterogeneity, and constraints of production networks. Beyond \textit{ReGeNet}, this work contributes a failure taxonomy of multimodal \ac{LLM}-based \ac{IBN} systems and a scalability characterization of \ac{LLM} topology understanding across networks of up to 750 nodes.
\end{abstract}

\section{Introduction}\label{sec:intro}

Communication network engineering in enterprise environments is complex, time-consuming, and error-prone, relying on large \ac{IT}, \ac{NetOps}, and network engineering teams to design topologies, map networks, and configure components according to established policies and security best practices~\cite{ifland2025genet}. Growing architectural diversity makes these tasks increasingly challenging~\cite{houidi2022neural}, requiring coordination across heterogeneous devices, systems, and standards.

To simplify these workflows, researchers have explored \acf{IBN}~\cite{leivadeas2022survey}, which bridges high-level intents and low-level implementations. Early \ac{IBN} frameworks relied on formal languages and optimization tools~\cite{beckett2016don}, offering determinism but limited flexibility. More recent approaches use \ac{NLP}-based tools such as \acp{LLM} to reduce this formality and assist engineers in generating device configurations and policies~\cite{wang2023making,jacobs2021hey}, inspired by programming co-pilots such as GitHub Copilot\footnote{https://github.com/features/copilot} and Amazon CodeWhisperer\footnote{https://aws.amazon.com/codewhisperer/}~\cite{guo2023configreco,zhao2023confpilot}.

Most network automation research, however, focuses on configuration synthesis, overlooking topology management. An exception is GeNet~\cite{ifland2025genet}, which integrates multimodal capabilities and \ac{VQA} techniques~\cite{ishmam2024image} to interpret topology sketches or images alongside device configurations, allowing topology images to serve directly as \ac{IBN} specifications~\cite{wang2024netconfeval}.

\acp{LLM} remain prone to errors and hallucinations, motivating verification mechanisms~\cite{wang2023making}. Traditional verification relies on formal analysis of configurations~\cite{mondal2023llms,mondal2025llm,tang2023lightyear}, which is rigid, resource-intensive, and insufficient for broader network engineering tasks~\cite{ifland2025genet}. Recent work has explored LLM-driven verification~\cite{wei2025leveraging}, but studies of verification quality, reasoning capabilities, and multimodal input remain sparse.

We introduce \textit{Reflective-GeNet (ReGeNet)}, a multi-agent co-pilot that extends \emph{GeNet} with a self-verification loop reducing \ac{LLM}-induced errors: one agent implements user intents while a second independently verifies outcomes and provides corrective feedback. \revised{Verification improves implementation quality for non-reasoning \acp{LLM} (+0.99 IIS points on average), enabling GeNet+V to perform comparably to the reasoning model baseline (GeNet+R). Reasoning-capable models see no significant additional benefit from verification, consistent with the broader self-correction literature~\cite{huang2024large}.}

\ac{ReGeNet} comprises two modules: \emph{Topology Understanding} (Section~\ref{sec:topology-understanding-method}) and \emph{Reflective Intent Implementation} (Section~\ref{sec:solver}). The former robustly interprets topology images of varying quality, including messy sketches and scanned diagrams. Across nine state-of-the-art \acp{LLM}, \textit{GPT-4.1-mini} and \textit{o4-mini} significantly outperform the others, with \textit{LLaMA-3.2-Vision-Instruct 11B} failing to generate valid topology JSON in 84\% of cases; \textit{GPT-4.1-mini} is also more token-efficient than \textit{o4-mini}.

The latter module shows that \revised{verification improves implementation quality for non-reasoning \acp{LLM} (\hyperref[sec:full_exp]{\textit{GeNet+V}}) by +0.99 IIS points on average, enabling it to match the reasoning model baseline (\hyperref[sec:full_exp]{\textit{GeNet+R}}); adding verification to a reasoning \ac{LLM} (\hyperref[sec:full_exp]{\textit{GeNet+RV}}) yields no significant further gain, with a point estimate of $-$0.19 IIS points, suggesting the two mechanisms provide partially overlapping benefits.} Non-reasoning \acp{LLM} maintain consistent quality across complexity levels, while reasoning LLMs do slightly better on easier scenarios; token usage scales with complexity and implementation time is similar across methods.

The example below illustrates the practical utility of \ac{ReGeNet}:

\begin{example}[Adding Local PCs]
\label{ex:scenario}
Management requests adding three workstations to a lab, but all ports on the corresponding switch are occupied. The engineer provides this intent, along with an image of the lab’s topology and the raw device configurations. The \emph{ReGeNet} implementation agent detects that the switch has no available ports and initially suggests connecting the new workstations directly to the department’s router. However, the \emph{ReGeNet} verification agent reviews this proposal, notes that it violates network design standards and best practices, and instead recommends adding another switch in a daisy chain and connecting the workstations to it.
The implementation agent accepts the feedback and generates the updated configurations and topology accordingly. 
\end{example}

In Section~\ref{sec:eval}, we evaluate \ac{ReGeNet} on a multimodal \ac{IBN} benchmark~\cite{ifland2025genet} based on Cisco certification exercises, spanning scenarios of varying complexity and input quality, examining topology understanding, token efficiency, and the combined impact of reasoning and verification on intent implementation quality. The results show that \ac{ReGeNet} improves reliability and handles both simple and complex intents effectively, highlighting its practical potential for enterprise network automation.

\subsection{Contributions} \label{sec:contribution}

This paper makes the following main contributions:

\begin{itemize}
    \item We present \ac{ReGeNet}, a multi-agent network copilot that employs a self-verification loop to \revised{reduce \ac{LLM}-induced semantic configuration errors in \ac{IBN}, providing an explicit, auditable correction trace.}
    
    \item We develop a \emph{Topology Understanding} module that \revised{converts network topology images---including low-quality sketches---into structured JSON representations using \acp{VLLM}. Two of nine benchmarked models (\textit{GPT-4.1-mini} and \textit{o4-mini}) achieve reliable performance; the remaining seven exhibit substantial failure rates.}
    
    \item We design a \emph{Reflective Intent Implementation} module in which implementation and verification agents iteratively refine configurations and topologies. This module introduces an automated verification mechanism that \revised{improves LLM-judged intent implementation quality, particularly for non-reasoning \acp{LLM} (+0.99 IIS points on average), and provides a verifier pass/fail signal that can support human oversight decisions in mission-critical deployments.}
    
    \item We benchmark nine state-of-the-art \acp{LLM} on topology understanding and evaluate the two best-performing models on intent implementation, revealing the impact of reasoning capabilities, token efficiency, and the contribution of \ac{ReGeNet}'s verification loop.
    
    \item We provide the first systematic study, to the best of our knowledge, of the interaction between \ac{LLM} reasoning and external verification in multimodal \ac{IBN}. Our results show that verification improves intent implementation quality for non-reasoning \acp{LLM}, enabling GeNet+V to match the reasoning model baseline (GeNet+R). \revised{Adding verification to a reasoning model (GeNet+RV) yields no significant further improvement, with a point estimate of $-$0.19 IIS points---a result consistent with the self-correction literature~\cite{huang2024large} and an informative negative finding in its own right.}
    
    \item We derive practical guidance for enterprise network engineers and system designers by showing that comparable performance can be achieved either through a lightweight non-reasoning model with verification or through a more expensive reasoning model without external verification, thereby enabling better trade-offs between quality, latency, and computational cost.
    
    \item We validate \ac{ReGeNet} on a multimodal \ac{IBN} benchmark with realistic Cisco-inspired scenarios, demonstrating its improved intent implementation quality, robustness to input variations, and practical efficiency for network copilot deployment compared to baselines. 
    Additionally, we provide a comprehensive engineering analysis of multimodal IBN systems, including LLM failure modes, verification-induced errors, and topology-understanding scalability \revised{up to a model-dependent reliable processing limit in the 150--500 node range, beyond which accuracy degrades sharply.}
\end{itemize}

The remainder of this paper is structured as follows. 
Section~\ref{sec:background_relate-work} provides background and reviews related work. 
The \ac{ReGeNet} framework and its core modules are introduced in Section~\ref{sec:genet-framework}. 
Section~\ref{sec:eval} presents our benchmark datasets, experimental setup, evaluation criteria, and results. 
Finally, in Section~\ref{sec:conclusion}, we summarize the main findings and outline directions for future work.

\section{Background and Related Work}
\label{sec:background_relate-work}

Modern enterprise networks require scalable, adaptive, and intelligent management. This section reviews prior work on automated configuration, LLM-driven networking, agentic frameworks, verification, and multimodal inference.

\subsection{Automated Network Configuration}
\label{sec:related:conf}

Manual network configuration is complex and error-prone. IBN automates this by translating high-level user intents into low-level device configurations~\cite{leivadeas2022survey}. \revised{The IETF has standardized core concepts and protocol foundations for this space: RFC~9315~\cite{rfc9315} defines intent-based networking concepts and terminology; RFC~6241~\cite{rfc6241} specifies the NETCONF protocol for configuration management, including \texttt{<validate>} and commit-confirmation operations; and RFC~7950~\cite{rfc7950} defines YANG, the data modeling language used in model-driven configuration. Production IBN systems typically land changes via NETCONF/YANG rather than raw CLI text, with atomicity, rollback, and configuration-drift detection built into the protocol layer. The present work generates CLI-level configurations---a practical choice for evaluation in GNS3-based environments---and engagement with the full NETCONF/YANG stack remains an important direction for future work (Section~\ref{sec:conclusion}).} Early systems relied on domain-specific languages~\cite{schneider2021snowcap}, while more recent approaches use structured data~\cite{ooi2023intent} or natural language inputs~\cite{wang2024netconfeval,khurana2023natural}.  
Intents have been mapped to configurations using compilers~\cite{beckett2016don}, deep learning models~\cite{guo2023configreco}, search-based algorithms~\cite{ooi2023intent}, and hybrid strategies.
For instance, Jacobs et al.~\cite{jacobs2021hey}, introduced LUMI, a system that utilizes machine learning and a dedicated compiler to translate operator intents into network configurations and deploy them.
Another example is NetMind~\cite{li2024netmind}, which combines Deep Reinforcement Learning (DRL) with \revised{Graph Convolutional Networks (GCNs)} to determine the function placement strategies in \revised{Radio Access Networks (RANs)} with diverse topologies, aiming at minimizing power consumption.
Each method balances transparency, flexibility, and computational cost differently. 

\subsection{\acp{LLM} and Network Configuration}
\label{sec:related:llm}
Transformer-based \acp{LLM}~\cite{shao2024survey} enable reasoning, code generation, and complex task execution. Researchers have explored the use of \acp{LLM} in networking to translate intents into device configurations~\cite{mekrache2024intent, chen2022software} and leverage APIs for complex tasks~\cite{huang2023large}. For instance, Zhao et al.~\cite{zhao2023confpilot} and Adanza et al.~\cite{adanza2025leveraging} combined LLMs with \ac{RAG} to generate and manage configurations from natural language intents. 
However, evaluations typically focus on minimal topologies or small configuration snippets. 
For instance, Lira et al. present LLM-NetCFG~\cite{lira2024large}, which introduces a privacy-aware architecture that employs a locally deployed \ac{LLM} to generate network configuration, but evaluates it solely on a four-router, two-host topology.
Donadel et al.~\cite{donadel2024llms} found that GPT-4 achieves the strongest results in network comprehension. 
Additionally, benchmarking frameworks such as NetConfEval~\cite{wang2024netconfeval}, which also explore the use of \acp{LLM} for network configuration generation, and NetLLMBench~\cite{aykurt2024netllmbench} have shown that GPT-4 performs best in \ac{IBN} tasks. 

Notable frameworks include Emergence~\cite{dzeparoska2025kpi}, which converts intents into MAPE-K~\cite{dzeparoska2021towards} policy trees, supporting monitoring but not direct configurations; ConfigTrans~\cite{zheng2024configtrans}, which combines \acp{LLM} with constraint-solving but faces scalability issues; Network CoPilot~\cite{han2025network}, which translates intents to device-specific configurations but struggles with heterogeneous deployments; and GeNet~\cite{ifland2025genet}, which leverages multimodal inputs, including topology images, to implement intents efficiently, but lacks mechanisms for output verification and refinement.

\subsection{Agentic-AI Frameworks}
\label{sec:related:agentic}

Agentic-AI integrates multiple specialized \ac{LLM} agents for coordination, complex reasoning, and closed-loop adaptation. Guo et al.~\cite{guo2024large} reviewed multi-agent \ac{LLM} systems, noting challenges in standardization and reproducibility.
In networking, Wang and Li~\cite{wang2025mnc} introduced a multi-agent \ac{LLM} framework for task decomposition and refinement, and Sun et al.~\cite{sun2025scnoc} developed \textit{SCNOC-Agentic} for satellite network control; however, these works lack multimodal inference and self-verification, which our work addresses (see also~\cite{li2022applications,amirkhani2022consensus,arzo2021multi}).

\subsection{Verification Using LLMs}
\label{sec:related:verification}
Verification is critical for \ac{LLM}-based \ac{IBN}, given \acp{LLM}' propensity for errors and hallucinations~\cite{wang2023making}. \revised{Traditional formal verification methods, e.g., Mondal et al.'s verifiers~\cite{mondal2025llm,mondal2023llms} and Batfish~\cite{beckett2017general}, offer rigorous, specification-driven guarantees and are not subject to the ambiguity that can affect natural-language descriptions. Their primary limitation for our setting is scope rather than rigor: they operate on fully specified, syntactically valid configurations and cannot reason over topology images, evaluate structural intent changes expressed in natural language, or handle intents that are partially underspecified---tasks that are central to multimodal IBN. LLM-based verification is therefore not a replacement for formal analysis but a complementary layer of intent-alignment checking. An important direction for future work, noted in our conclusions, is a hybrid architecture in which an LLM agent handles intent-level semantic validation while a formal analyzer provides configuration-level correctness guarantees downstream.}

\acp{LLM} offer a more adaptive alternative: Hu et al.'s \textit{QualityFlow}~\cite{hu2025qualityflow} uses an \ac{LLM} for program-synthesis quality checks, and Gu et al.~\cite{gu2024survey} survey \textit{LLM-as-a-judge} methods, noting limited transparency. In networking, Wei et al.~\cite{wei2025leveraging} used \acp{LLM} for semantic configuration verification; however, combining \ac{LLM} verification with reasoning in multimodal enterprise \ac{IBN} remains largely unstudied.

\revised{A line of work on LLM self-correction is directly relevant to our design. Huang et al.~\cite{huang2024large} show that LLMs cannot reliably improve their own outputs through intrinsic self-critique without external signal; Shinn et al.~\cite{shinn2023reflexion} and Madaan et al.~\cite{madaan2023self} propose reinforcement- and feedback-based refinement loops that rely on external evaluation signals to drive improvement; and Stechly et al.~\cite{stechly2023gpt} demonstrate that iterative prompting without grounded feedback often fails to correct reasoning errors. Our GeNet+RV result---where same-model verification does not significantly improve over the unverified reasoning baseline---is consistent with these findings, and we explicitly frame it as a replication of this negative result in the multimodal IBN domain. Importantly, these works also suggest that \emph{cross-model} or \emph{externally grounded} verification loops are more likely to provide genuine benefit, motivating future work on heterogeneous-agent verification for IBN.}

\subsection{Multimodal Inference and VQA}
\label{sec:related:visual}
Multimodal inference extends \ac{NLP} with visual information, exemplified by \ac{VQA}, answering questions grounded in images. Early convolutional/recurrent \ac{VQA} models~\cite{sharma2021survey} generalized poorly; vision-capable \acp{LLM} now enable end-to-end multimodal inference~\cite{liu2024visual}, e.g., diagram interpretation, evaluated via benchmarks like~\cite{babkin2023bizgraphqa}.

In networking, our prior work~\cite{ifland2025genet} used \ac{VQA} to describe topology images for multimodal \ac{IBN}; its unstructured outputs motivated the structured \ac{LLM} outputs used here.

In summary, prior \ac{IBN}, LLM-configuration, agentic-AI, verification, and multimodal-inference work remains limited: \acp{LLM} still hallucinate, verification centers on formal configuration checks, topology understanding struggles with low-quality diagrams, and \ac{LLM}-based topology verification is rarely integrated. \emph{ReGeNet} addresses these gaps by combining multimodal topology understanding, multi-agent inference, and \ac{LLM} verification, leveraging both reasoning and non-reasoning \acp{LLM} to reduce errors and improve intent implementation accuracy in realistic enterprise scenarios.

We extend our previous study~\cite{ifland2025genet} as follows:
\begin{itemize}
    \item A self-verification loop extending GeNet, improving reliability and reducing human intervention (\Cref{sec:solver}).
    \item Improved topology understanding via structured \ac{LLM} responses (\Cref{sec:topology-understanding-method}).
    \item A benchmark of nine state-of-the-art \acp{LLM} on topology understanding (Section~\ref{sec: vision_bench}).
    \item An intent-implementation benchmark of the two leading \acp{LLM}, comparing reasoning versus non-reasoning models (Section~\ref{sec:full_exp}).
    \item An analysis of the token usage-quality tradeoff (Sections~\ref{sec:vision_bench_results} and~\ref{sec:survey_res}).
    \item A domain expert survey on scenario complexity, aligned with system performance (\Cref{sec:survey_res}).
\end{itemize}


\section{The \ac{ReGeNet} Framework} \label{sec:genet-framework}

This section describes \ac{ReGeNet}, an agentic framework for \ac{IBN} designed to mitigate \ac{LLM}-induced errors and hallucinations through a built-in self-verification loop. 
Provided with a natural language intent and the current network state (topology and device configurations), \ac{ReGeNet} produces updated specifications that implement the intent while maintaining correctness.  

\begin{figure}[b]
    \centering
    \includegraphics[width=0.35\textwidth]{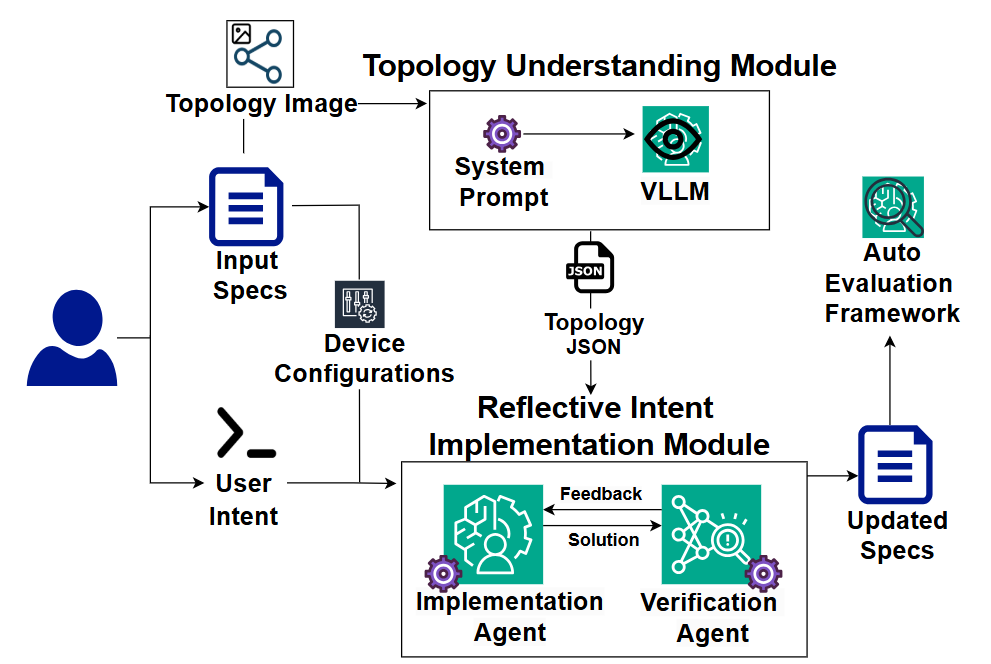}
    \caption{Overview of \ac{ReGeNet}: topology interpretation, iterative intent implementation with self-verification, and automatic evaluation of the updated network specification.}
    \label{fig:architecture}
\end{figure}

As shown in \Cref{fig:architecture}, \ac{ReGeNet} has a modular design, motivated by the observation that \acp{LLM} show greater reliability when complex tasks are broken down into smaller, well-defined components~\cite{wang2024netconfeval}. 
The framework consists of three core modules: 

\begin{itemize}
    \item \emph{Topology Understanding}, which converts a network topology image into a structured JSON representation;  
    \item \emph{Reflective Intent Implementation}, in which an implementation agent and a verification agent iteratively refine topology and configuration updates until the intent is satisfied;  
    \item \emph{Automatic Evaluation Framework}, which assesses the quality of the resulting solutions utilizing an \ac{LLM}.  
\end{itemize}
This architecture advances prior LLM-based IBN approaches by embedding reflection and verification directly into the workflow. The result is a more reliable and adaptable framework for intent implementation. 

\ac{ReGeNet} is a direct continuation of our earlier work on \textit{GeNet}~\cite{ifland2025genet}, which demonstrated the potential of multimodal copilots for interpreting both textual and visual representations of networks. While \textit{GeNet} focused on topology understanding and multimodal intent implementation, it lacked mechanisms for verifying and refining outputs. \ac{ReGeNet} addresses this limitation by integrating a self-verification loop, thereby providing a more comprehensive and trustworthy \ac{IBN} framework.  
In the following subsections, we describe each module in detail and present the research questions that guide our evaluation.
 
\subsection{Topology Understanding Module}
\label{sec:topology-understanding-method}
The \emph{Topology Understanding} module converts a network topology image into a structured JSON representation, leveraging \acp{LLM}' \ac{VQA} capabilities under carefully designed output constraints. The model receives the image alongside a detailed system prompt (\Cref{fig:vision_prompt}, shown as \emph{System Prompt} in \Cref{fig:architecture}).
\begin{figure}[h]
    \centering
    \scalebox{0.85}{%
    \begin{minipage}{1.15\linewidth}
\begin{lstlisting}[
        breaklines,
        numbers=left,
        frame=single,
        basicstyle=\ttfamily\tiny,
        showspaces=false
    ]
As a network topology analyst, you are given an image of a network topology diagram.
Analyze it carefully and return **only** a single JSON object in the exact format below, without extra text.

  {
    "nodes": [
      {
        "label": "<NODE_NAME>", //  Avoid adding component type to the name of the component i.e. router R1 should be called "R1" rather than "Router R1"
        "icon": "<one of: pc | cloud | router | ethernet_switch | ids>",
        // If icons are not illustrated in the diagram, use an empty string ("") for the icon field.
        // If the icon's name differs, map it to the closest option above e.g. "Desktop computer symbol" -> "pc".      
      }
      // ...repeat for every device in the topology (routers, firewalls, switches, PCs, etc.)
    ],

    "links": [
      ["<source_node_label>", "<destination_node_label>", "<source_interface>", "<destination_interface>"]
      // If interfaces are not illustrated in the diagram, use an empty string ("") for both interface fields.
      // ...repeat for every edge/connection in the topology
    ]
  }
\end{lstlisting}
    \end{minipage}}
    \caption{The \emph{Topology Understanding} system prompt.}
    \label{fig:vision_prompt}
\end{figure}
Acting as a network topology analyst, the \ac{LLM} must output a predefined JSON schema of nodes and links: each node has a label and, if available, an icon from a restricted set (empty string otherwise, to limit hallucination); each link is a source-target tuple with interfaces, similarly defaulting to empty strings when unavailable. The prompt enforces JSON-only responses with the schema repeated per node and link, reducing omissions and ensuring uniform output; these constraints were iteratively refined in preliminary experiments to improve consistency and robustness.

Building on~\cite{ifland2025genet}, this module extends GeNet's multimodal pipeline with stricter output constraints and schema enforcement, yielding a more reliable, structured translation of visual topologies that underpins the subsequent reflective intent implementation module.

\revised{Several corner cases challenge the robustness of the Topology Understanding module. Low-resolution images reduce the effective pixels per node available to the vision model's tokenizer, making device labels and link endpoints difficult to resolve; this effect compounds as topology size grows, and is a contributing factor to the phase transition observed in Section~\ref{sec:scalability}. Overlapping edges---present by design in our \emph{Messy Layout} variants---similarly degrade link detection accuracy, as the model must disambiguate visually coincident paths. When the source diagram file is available (e.g., a GNS3 project file or a draw.io XML export), a deterministic parser would bypass these failure modes entirely and is preferable; the LLM vision pipeline is intended for cases where only a rasterized image is available, such as scanned hand-drawn sketches or photographed whiteboards.}

\subsection{Reflective Intent Implementation Module}
\label{sec:solver}

The \emph{Reflective Intent Implementation} module updates network specifications to align with user intents while minimizing \ac{LLM}-induced errors and hallucinations. It builds on Microsoft’s Autogen framework~\cite{wu2024autogen}, which enables multi-agent collaboration, and leverages OpenAI’s Assistants API\footnote{https://platform.openai.com/docs/assistants/overview?context=with-streaming} for file exchange and state management during interactions.

This module consists of two agents: the \emph{implementation agent}, which applies user-specified changes to the topology and device configurations, and the \emph{verification agent}, which validates the updates and provides corrective feedback. The agents iterate until consensus is reached, ensuring correctness. In addition, the implementation agent generates human-readable explanations of the changes applied, providing both information to the user and transparency.

As in the \emph{Topology Understanding} module, both agents are initialized with detailed system prompts that specify their roles, responsibilities, and expected outputs. These instructions were refined in preliminary experiments to achieve stable and reproducible performance.  
The implementation code and prompts are available in our Git repository.\footnote{\url{https://github.com/networkcopilot/ReGeNet/tree/main/Code/Reflective\%20Intent\%20Implementation\%20Module}} 

\subsubsection{Implementation Agent}

The implementation agent functions as an expert in network topology and configuration. Its task is to faithfully execute user intents by updating the JSON topology and modifying device configurations as required. Because early experiments showed frequent partial or misinterpreted updates, the agent is provided with structured guidelines to ensure completeness. These guidelines cover tasks such as creating new nodes or links, editing existing configurations, and preserving schema consistency.  

The agent submits its modifications to the verification agent for review and iteratively incorporates feedback until it obtains full approval. Afterward, it generates a structured summary of all changes and produces an updated file list. Although the framework itself was not adapted to any specific configuration language, our empirical evaluation is limited to Cisco-inspired scenarios; validating performance on additional vendors and configuration syntaxes remains an important direction for future work.

\subsubsection{Verification Agent} \label{sec:method_verifier}
The verification agent aims to improve the likelihood that updates produced by the implementation agent are both correct and aligned with the user’s intent, while also preventing unintended side effects. 
Its instructions, adapted from our automatic evaluation framework (\Cref{sec:llm_eval}), remain generic rather than scenario-specific, making the agent broadly applicable.  

This agent reviews the original and updated files, checks compliance with the user’s intent, and records all approved files. It also enforces strict naming conventions to maintain consistency across iterations. During operation, it reloads the verified file list, only approves new updates if fully correct, and re-examines previously approved files if subsequent changes affect them. 

\subsection{Automatic \ac{LLM}-Based Evaluation}
\label{sec:llm_eval}

For \ac{ReGeNet} to function as a reliable evaluator, its \ac{LLM}-based evaluation must align closely with human judgments. We therefore adopt GeNet’s automatic evaluation framework for \ac{IBN} intent implementations~\cite{ifland2025genet}, which demonstrated high correlation with expert assessments. This framework rests on the hypothesis that evaluating implementations against predefined scoring keys is simpler for \acp{LLM} than producing the implementations themselves.  

The evaluation module uses the Assistants API, where \revised{the evaluation \ac{LLM}---\textit{GPT-4o} (accessed via the OpenAI Assistants API at temperature 0)---}receives (i) explicit evaluation instructions, (ii) the task-specific scoring keys (\Cref{sec:fst-scrn}), and (iii) the network topology and device configurations before and after intent implementation. \revised{We note that temperature 0 reduces but does not eliminate non-determinism in the OpenAI API; Table~\ref{tab:verification_degradation_taxonomy} shows that 25\% of verification-induced degradations are classified as Re-evaluation Inconsistency, reflecting judge variability even at temperature 0. This noise level should be interpreted alongside the +0.99 IIS effect being defended, and constitutes a limitation of LLM-based evaluation.} To ensure reliability, the evaluator receives the ground-truth topology (not the one generated by the \emph{Topology Understanding} module) as input for scoring purposes only. Importantly, the \acf{IIS} is computed based on the outputs of the full pipeline, where the Implementation module operates on the topology produced by the Topology Understanding module. Therefore, any errors introduced in topology understanding do propagate into the implementation stage and are reflected in the \ac{IIS}. 
The use of ground-truth topology at the evaluation stage is solely to ensure accurate and consistent assessment of the final outcome, rather than to bypass upstream errors.
Importantly, the evaluation is conducted independently, without access to the \textit{Intent Implementation module} agents’ explanations. This design ensures that the assessments are unbiased and based solely on the final outcomes. The complete instructions~\footnote{\url{https://github.com/networkcopilot/ReGeNet/tree/main/Code/Automatic\%20Evaluation\%20Framework}} and scoring keys~\footnote{https://github.com/networkcopilot/ICDCS25-GeNet/tree/main/Dataset} are available in our Git repository.

\subsection{Research Questions}\label{sec:rq}

Our evaluation is guided by the following research questions, which were formulated in response to the limitations of prior work. Prior work on IBN has largely overlooked topology-level inference, lacked practical verification mechanisms, and rarely examined efficiency or complexity factors. To address these gaps, we structure our investigation around the following five questions:

\begin{resq}\label{rq:images}
How robustly can \acp{VLLM} interpret network diagrams of varying quality and layout complexity? (\Cref{sec:vision_bench_results})  

\textit{Motivation:} Previously proposed frameworks focused mainly on textual configuration synthesis, leaving the interpretation of messy or imperfect topology diagrams underexplored.
\end{resq} 

\begin{resq} \label{rq: verification_effectiveness}
To what extent does a multi-agent self-verification loop improve the quality of intent implementations in multimodal \ac{IBN}? (\Cref{sec:implementation_res}) 

\textit{Motivation:} Existing verification techniques emphasize formal configuration analysis, which is rigid and impractical for holistic frameworks that must also handle topologies.
\end{resq}

\begin{resq} \label{rq: verification_vs_reasoning}
How does the reasoning ability of \acp{LLM} influence the utility of external verification mechanisms in multi-agent systems? (\Cref{sec:implementation_res})  

\textit{Motivation:} With reasoning-enhanced \acp{LLM}, it remains unclear whether reflective verification provides substantial benefits or whether reasoning suffices.
\end{resq}

\begin{resq}\label{rq:questionnaire}
How does the perceived complexity of intents, as judged by domain experts, align with the quality of topology understanding and intent implementation achieved by an \ac{LLM}-based networking co-pilot? (\Cref{sec:survey_res})  

\textit{Motivation:} Prior studies rarely evaluated how system performance corresponds to expert expectations, leaving the question of real-world usability and trustworthiness open.
\end{resq}

\begin{resq}\label{rq:efficiency}
What are \ac{ReGeNet}'s efficiency trade-offs, in terms of token usage, execution time, and practicality for enterprise-scale deployment? (\Cref{sec:vision_bench_results} and \Cref{sec:survey_res})  

\textit{Motivation:} While previous studies have focused mainly on accuracy and correctness, the real-world adoption of multimodal \ac{IBN} systems depends on their operational efficiency. Understanding how computational cost and latency evolve across different \ac{LLM} architectures and input variations is essential to designing deployable and cost-effective network co-pilots.
\end{resq}
These research questions move from examining multimodal robustness (RQ1), to addressing verification and reasoning (RQ2, RQ3), aligning technical outcomes with expert judgment (RQ4), and finally, exploring efficiency and scalability (RQ5). This progression allows us to systematically address prior work's open challenges and demonstrate how \ac{ReGeNet} advances the state of the art in reflective, multimodal \ac{IBN}.

\section{Experiments}\label{sec:eval}

This section describes the experiments conducted to investigate the research questions presented in Section~\ref{sec:rq}. Our goal is to assess the performance and capabilities of the \ac{ReGeNet} framework, as well as the factors that influence its effectiveness.  

We begin with a description of the multimodal \ac{IBN} benchmark dataset introduced in our previous study~\cite{ifland2025genet}, which includes scenarios derived from Cisco certification exercises and multiple variants of topology images (Section~\ref{sec:data}). We then discuss the experimental setup (Section~\ref{sec:exp-setup}) and criteria used to evaluate the quality of both topology interpretation and intent implementation (Section~\ref{sec:quality}). Next, we present our domain expert survey, which is used to assess the scenario complexity (Section \ref{sec:survey}). Finally, we provide and discuss our experimental results (Section~\ref{sec:results}).

\subsection{Datasets}\label{sec:data}

We constructed our dataset by adapting training scenarios from GNS3Vault,\footnote{https://gns3vault.com} an online repository that provides downloadable GNS3 labs, topologies, and configurations to help users practice and learn real-world network setups using Cisco and other vendor devices. In collaboration with IT experts, we designed 10 scenarios, each comprising a natural language intent and detailed network specifications, including device configurations and topology images. Five scenarios focus on topology modifications, while the other five target configuration updates (Table~\ref{tab:scenario_categories}). The two groups are respectively referred to as \emph{topology} and \emph{configuration} scenarios. This dataset is referred to as the \emph{Original} dataset and is publicly available in GeNet's Git repository.\footnote{https://github.com/networkcopilot/ICDCS25-GeNet/tree/main/Dataset} To ensure realistic specifications, all scenarios were initially implemented in a network emulator, from which we extracted topology images and device configurations.

Although derived from certification-oriented training material, the selected GNS3 Vault scenarios capture representative network engineering tasks commonly encountered in enterprise environments, including topology modification, routing, access control, and service configuration. However, these scenarios remain smaller, more structured, and less heterogeneous than production enterprise networks. Consequently, our evaluation should be interpreted as demonstrating \textit{ReGeNet}'s effectiveness on representative network engineering tasks rather than as a complete validation on large-scale operational environments.

\begin{table}[ht]
\centering
\renewcommand{\arraystretch}{1.5}
\begin{tabular}{|m{3.8cm}|m{4.5cm}|}
\hline
\rowcolor{gray!20}
\textbf{Configuration Scenarios} & \textbf{Topology Scenarios} \\ \hline
IP Traffic Export (M) & Adding DMZ (M) \\ \hline
Basic Zone-Based Firewall (H) & Adding DRA (H) \\ \hline
Role-Based CLI Access (E) & Internet Connectivity (M) \\ \hline
Time-Based Access List (E) & Adding Local PCs (E) \\ \hline
IOS Transparent Firewall (M) & Adding Communication Servers (E) \\ \hline
\end{tabular}
\caption{Scenarios grouped by type and categorized by complexity level (E = Easy, M = Medium, H = Hard), based on a survey (Section~\ref{sec:survey}).}
\label{tab:scenario_categories}
\end{table}

\begin{figure*}[t]
    \centering
    \revised{\includegraphics[width=0.9\linewidth]{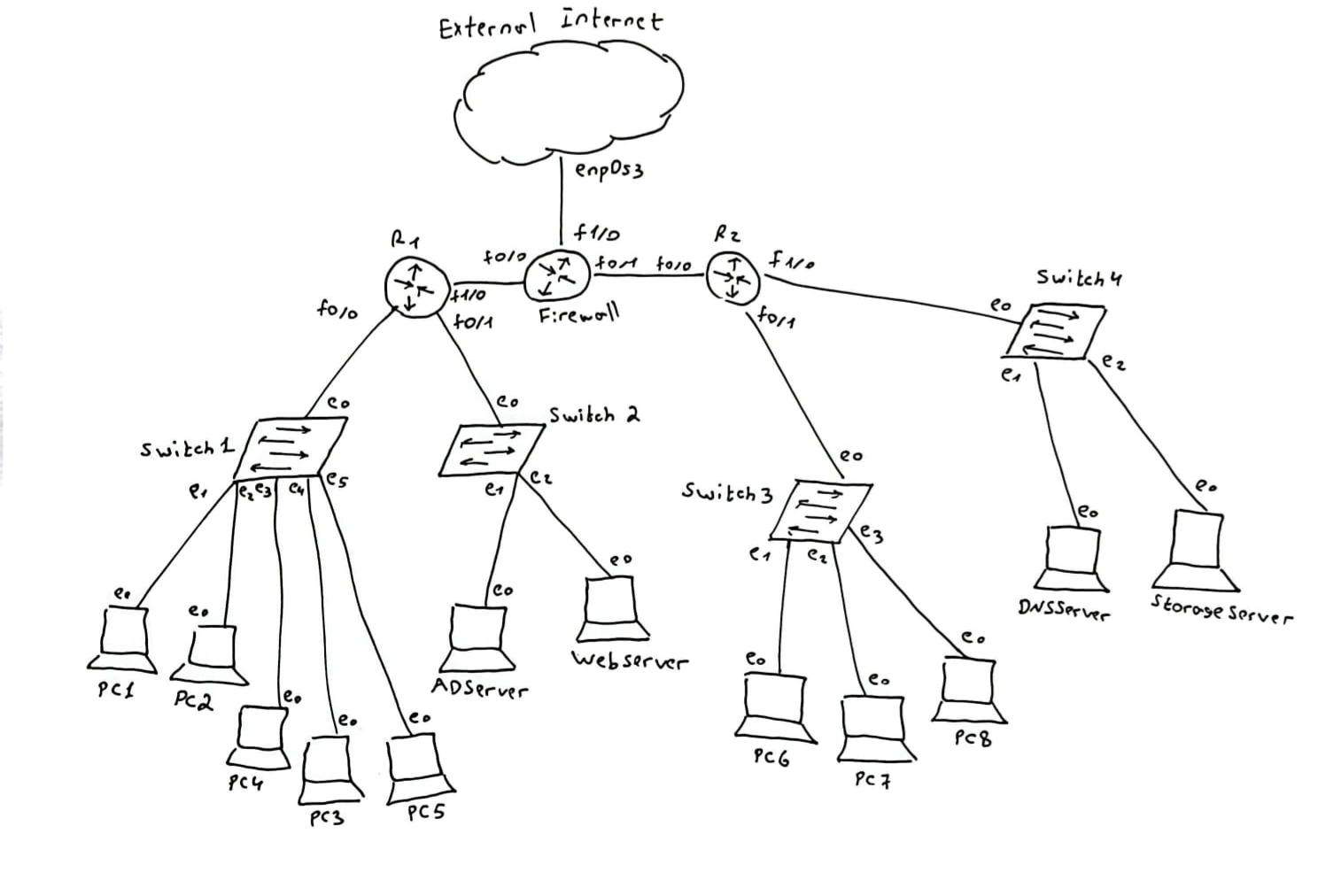}}
    \caption{Normal - paper sketch example.}
    \label{fig:exmp_normal_paper}
\end{figure*}

\begin{figure*}[t]
    \centering
    \revised{\includegraphics[width=0.9\linewidth]{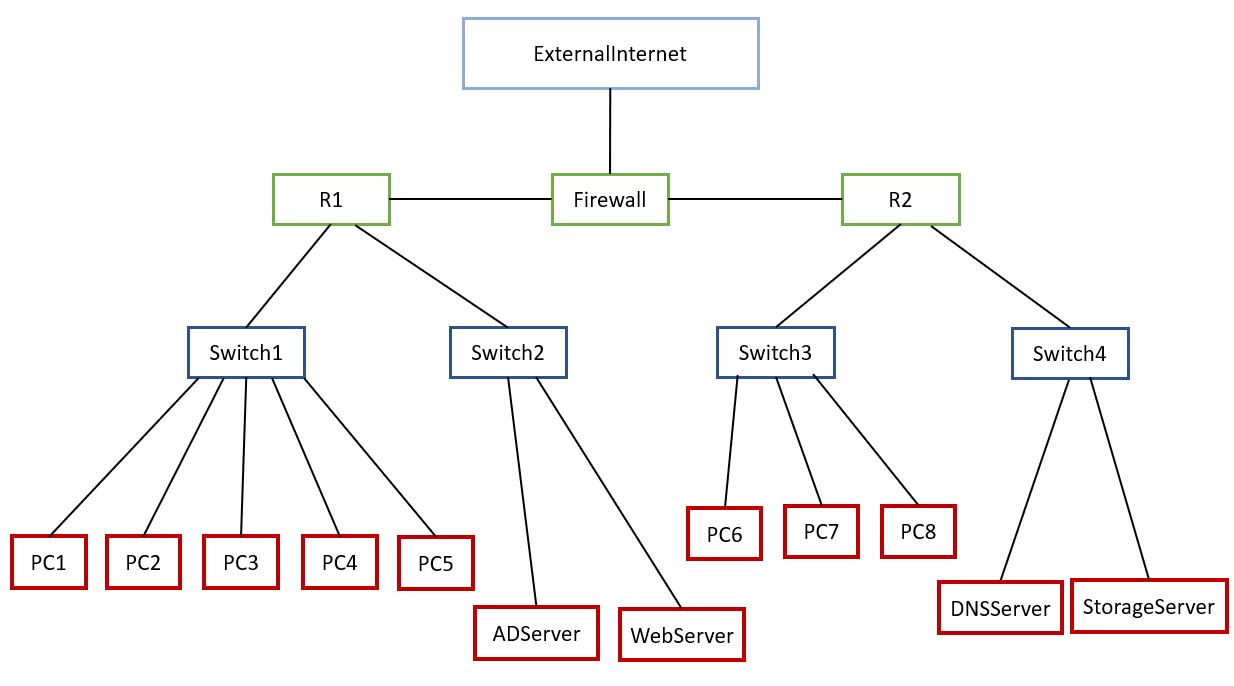}}\caption{No Labels on Edges - PowerPoint example.}
    \label{fig:exmp_no_labels_pp}
\end{figure*}

\begin{figure*}[t]
    \centering
    \revised{\includegraphics[width=0.9\linewidth]{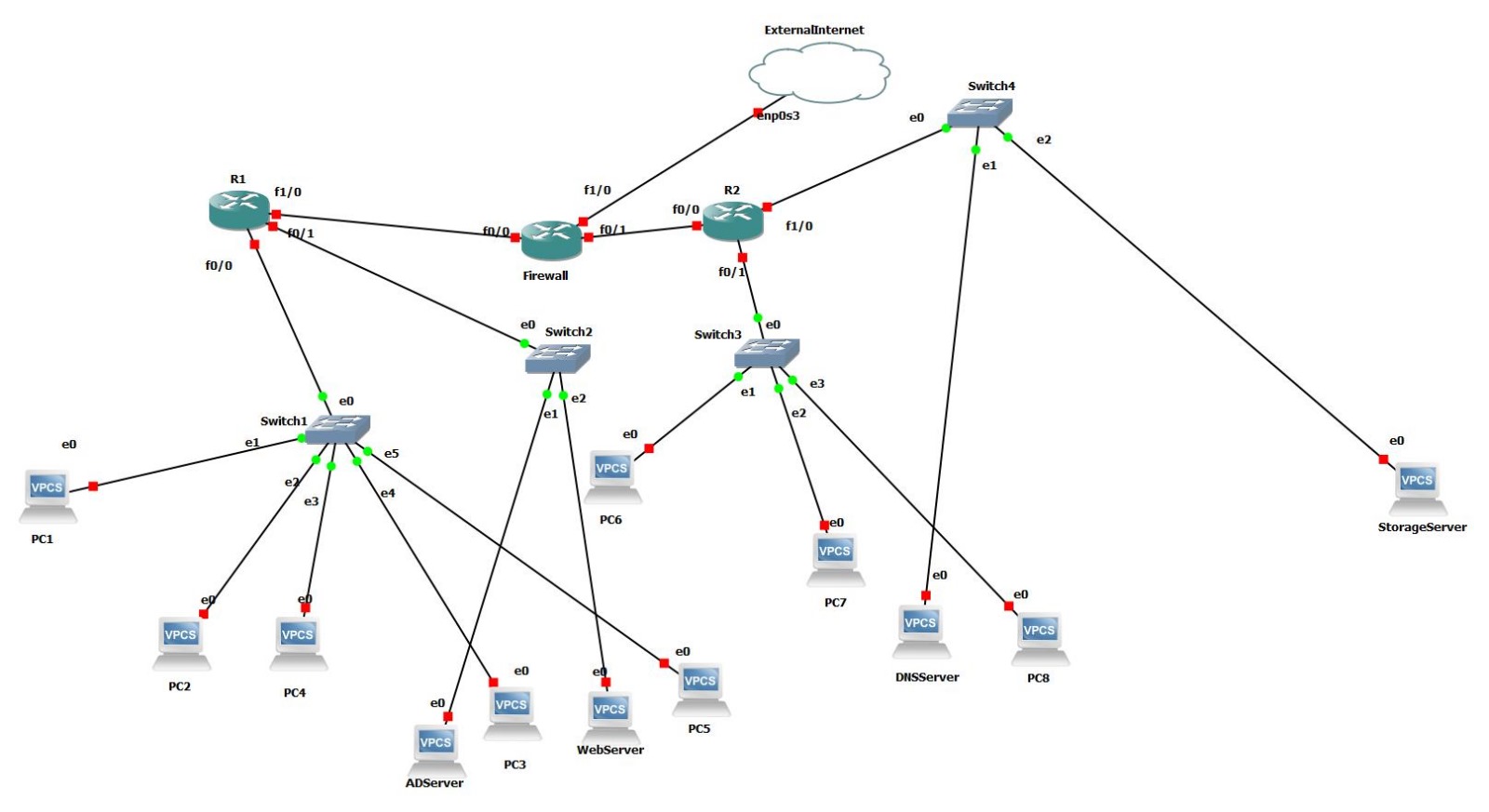}}\caption{Messy Layout - GNS3 example. ~\cite{ifland2025genet}}
    \label{fig:exmp_messy_gns}
\end{figure*}

To rigorously evaluate \acp{LLM} on multimodal input, we further extended the dataset with nine topology image variants per scenario, covering three visualization formats: (1) GNS3, (2) PowerPoint (without component icons), and (3) hand-drawn sketches. For each format, three types of images were created: (a) \emph{Normal}, (b) \emph{No Labels on Edges}, and (c) \emph{Messy Layout}. The \emph{Normal} variant is a clean, balanced topology (Figure \ref{fig:exmp_normal_paper}), while the \emph{No Labels} variant omits edge interface labels (Figure \ref{fig:exmp_no_labels_pp}). The \emph{Messy Layout} variant (Figure \ref{fig:exmp_messy_gns}) deliberately introduces visual challenges such as edge congestion, overlaps, and asymmetric layouts while preserving human readability. This augmented collection is referred to as the \emph{Vision-Extended} dataset.

Finally, to evaluate scalability, we constructed a topology dataset derived from a realistic enterprise bank network comprising 88 nodes. 
It was selected because it provides a representative enterprise architecture with diverse device types and is organized hierarchically. 
Smaller topologies (10-75 nodes) were generated by extracting connected Breadth First Search (BFS) subgraphs from the original topology, while larger topologies (100–750 nodes) were synthesized using a hierarchical topology generator we implemented\footnote{https://github.com/networkcopilot/ReGeNet/tree/main/Code/Topology\%20Scaler} that preserves the same architectural principles. 
This process resulted in 14 connected topologies ranging from 10 to 750 nodes, each accompanied by a rendered network topology image and a corresponding ground-truth JSON representation for evaluation.
This controlled process isolates the effect of topology scale while preserving consistent structural characteristics across all evaluated network sizes. 
This augmented collection of topologies is referred to as the \emph{Topology@Scale} dataset.

Together, the \emph{Original}, \emph{Vision-Extended} and \emph{Topology@Scale} datasets form a diverse, carefully designed benchmark that enables systematic analysis of the effects of scenario type, intent complexity, topology scale, and visualization quality on \ac{ReGeNet}, addressing the research questions in Section~\ref{sec:rq}.

\subsection{Experimental Setup}\label{sec:exp-setup}



\revised{\clearpage}
\begin{table}[]
\centering
\caption{Experimental setup: initial experiment (Init. Expt.), automated LLM-based evaluation (Auto. Eval.), \emph{Topology Understanding} benchmark, full experiment (Full Expt.), and \emph{Topology Understanding} scalability benchmark (Topology@Scale Benchmark).}
\label{tab:exp_setup}
\resizebox{\linewidth}{!}{%
\begin{tabular}{|l|c|c|c|c|c|}
\hline
\rowcolor[HTML]{C0C0C0}
 & \shortstack{\textbf{Init.}\\\textbf{Expt.}} & \shortstack{\textbf{Auto}\\\textbf{Eval}} & \shortstack{\textbf{Topology Understand.}\\\textbf{Benchmark}} & \shortstack{\textbf{Full}\\\textbf{Expt.}} & \shortstack{\textbf{Topology@Scale}\\\textbf{Benchmark}} \\ \hline
Dataset        & Original & Original & \shortstack{Vision-\\Extended} & \shortstack{Vision-\\Extended} & Topology@Scale \\ \hline
Models         & 1   & 1   & 9    & 2   & 3   \\ \hline
Temperatures   & 5   & 1   & 1    & 1   & 1   \\ \hline
Scenarios      & 10  & 10  & 10   & 10  & N/A \\ \hline
Image Variants & 1   & 1   & 9    & 9   & 14  \\ \hline
Test Cases     & 50  & 50  & 810  & 180 & 42  \\ \hline
Iterations     & 5   & 5   & 5    & 5   & 10  \\ \hline
Solutions      & 250 & 250 & 4050 & 900 & 420 \\ \hline
\end{tabular}%
}
\end{table}

\subsubsection{Initial Experiment} \label{sec:init_exp}

To examine the effect of stochasticity on \ac{ReGeNet}, we performed a preliminary temperature-sensitivity experiment using the GeNet framework~\cite{ifland2025genet}. We evaluated five temperature settings, \([0,0.1,0.5,0.9,1]\), on the Original dataset, executing each scenario five times per temperature. This resulted in 250 \textit{GeNet} solutions across 50 test cases (Table~\ref{tab:exp_setup}). The experiment, conducted using GPT-4 Turbo, indicated that temperatures in the range $[0,1]$ have only a negligible effect on performance, while higher temperatures often produced unstable outputs. The detailed results are therefore reported in the appendix, and we use the default temperature of 1 in the remainder of the paper.

\subsubsection{Automatic \ac{LLM}-Based Evaluation} \label{sec:exp_LLM_eval}

Manual evaluation of the initial experiment's solutions, using predefined scoring criteria (\Cref{sec:fst-scrn}), provided a reliable measure of performance but was time-consuming and limited in scalability. To overcome these issues, we developed an automated evaluation framework leveraging \acp{LLM}~\cite{ifland2025genet}.

The evaluating LLM was configured with a temperature of 0 to ensure deterministic scoring according to the predefined keys. The evaluation process involves providing the LLM with explicit instructions, the scoring keys, and the initial and updated network specifications. By comparing the initial specifications with the outputs generated by \ac{ReGeNet}, the model produces a quantitative assessment that strongly correlates with human evaluations. 
This approach allowed us to scale experiments across the expanded Vision-Extended dataset efficiently, while maintaining rigorous and reproducible evaluation of intent implementations.

\subsubsection{Topology Understanding Benchmark} \label{sec: vision_bench}
We first evaluate nine state-of-the-art \acp{LLM} on topology interpretation using our Vision-Extended dataset (\Cref{sec:data}). The models include five proprietary and four open-source models from three different providers, all equipped with vision capabilities to handle \ac{VQA} tasks. Each topology image variant for every scenario was processed five times by each model, resulting in a total of 4,050 topology JSON representations across 810 test cases (see Table~\ref{tab:exp_setup}). For brevity, we refer to this experiment as the \emph{Vision Bench} experiment.

\subsubsection{Full Experiment} \label{sec:full_exp} 
In the full experiment, we leverage the Vision-Extended dataset, which uses nine image variants per scenario. 
Based on insights from the initial experiment's results (Section~\ref{sec:initial_results}), we set the temperature at 1, which is the default for the Assistants API. To manage both coverage and cost, we chose two \acp{LLM} from a list of nine, based on earlier research~\cite{wang2024netconfeval,aykurt2024netllmbench,donadel2024llms,miao2023empirical} and the results from the Vision Bench experiment (\Cref{sec:vision_bench_results}). One of the selected models has the new LLM reasoning ability. We processed each version of the topology image five times with \emph{ReGeNet} using both models. 
This resulted in 900 \emph{ReGeNet solutions} across 180 test cases (Table~\ref{tab:exp_setup}).

To distinguish between the different framework variants, we use the following notation. The symbol ``+'' denotes our updated implementation framework relative to the original \textit{GeNet} baseline. The suffix ``V'' indicates that the proposed reflective verification loop is enabled, while ``R'' indicates the use of a reasoning-capable \ac{LLM}. Thus, \emph{GeNet+V} and \emph{GeNet+RV} correspond to the proposed \textit{ReGeNet} framework instantiated with non-reasoning and reasoning \acp{LLM}, respectively. In contrast, \emph{GeNet+} and \emph{GeNet+R} disable the verification mechanism and are included only as ablation baselines to isolate the contribution of the proposed reflective loop.

To evaluate the contribution of the proposed reflective mechanism, we compare two \textit{ReGeNet} variants with two ablation baselines:

\begin{itemize}
    \item \emph{GeNet+V}: the proposed \textit{ReGeNet} framework utilizing a non-reasoning \ac{LLM} (\textit{GPT-4.1-mini}) together with the reflective verification loop.
    
    \item \emph{GeNet+RV}: the proposed \textit{ReGeNet} framework utilizing a reasoning \ac{LLM} (\textit{o4-mini}) together with the reflective verification loop.
    
    \item \emph{GeNet+}: an ablation baseline identical to \emph{GeNet+V}, but without the verification loop.
    
    \item \emph{GeNet+R}: an ablation baseline identical to \emph{GeNet+RV}, but without the verification loop.
    
    \item \emph{GeNet}~\cite{ifland2025genet}: the original framework from prior work.
\end{itemize}

The ablation variants \emph{GeNet+} and \emph{GeNet+R} allow us to isolate the effect of the proposed verification mechanism for non-reasoning and reasoning \acp{LLM}, respectively.

\begin{table}[ht]
\centering
\footnotesize
\renewcommand{\arraystretch}{1.05}
\setlength{\tabcolsep}{3pt}

\begin{tabular}{|p{1.25cm}|p{1.35cm}|p{1.55cm}|p{3.8cm}|}
\hline
\rowcolor{gray!20}
\textbf{Method} &
\textbf{Model} &
\textbf{Verification} &
\textbf{Purpose / RQs} \\ \hline

GeNet~\cite{ifland2025genet}
& GPT-4 Turbo
& \revised{False}
& Original baseline
\\ \hline

GeNet+
& GPT-4.1-mini
& \revised{False}
& Non-reasoning ablation (RQ2)
\\ \hline

GeNet+V
& GPT-4.1-mini
& \revised{True}
& Proposed ReGeNet (RQ2, RQ4)
\\ \hline

GeNet+R
& o4-mini
& \revised{False}
& Reasoning ablation (RQ3)
\\ \hline

GeNet+RV
& o4-mini
& \revised{True}
& Proposed ReGeNet (RQ2--RQ4)
\\ \hline

\end{tabular}

\caption{\centering Methods and baselines used in the experiments.}
\label{tab:methods_and_baselines}
\end{table}
\Cref{tab:methods_and_baselines} summarizes the methods, clarifying which variants correspond to the proposed \textit{ReGeNet} framework and which serve as ablation baselines.

\subsubsection{Topology Understanding Scalability} \label{sec:scale_experimental}
To evaluate the scalability of the \emph{Topology Understanding} Module, we leveraged the \emph{Topology@Scale} dataset (\Cref{sec:data}), which contains 14 network topologies ranging from 10 to 750 nodes while preserving a consistent hierarchical enterprise structure. Based on the results of the Vision Bench experiment (\Cref{sec:vision_bench_results}), we selected the three highest-performing \acp{LLM}: GPT-4.1-mini, o4-mini, and Gemini-2.5-Flash. The temperature was set to 1, consistent with the full experiment. Each topology was processed independently 10 times with each model, resulting in 420 topology understanding runs across 42 test cases (Table~\ref{tab:exp_setup}).

\subsubsection{Comparison with SOTA Frameworks} \label{sec:experimenta_compare}
To compare \emph{ReGeNet} against existing \ac{LLM}-based \ac{IBN} frameworks, we selected \emph{NetConfEval}~\cite{wang2024netconfeval} and \emph{LLM-NetCFG}~\cite{lira2024large}, two representative state-of-the-art frameworks for \ac{LLM}-based network configuration generation.
Since neither framework includes topology understanding capabilities, in this experiment, all methods (including \emph{ReGeNet}) were provided with the same ground-truth JSON representation of the network topologies, thereby isolating the intent implementation task.
This, in turn, made our topology image variants dataset inapplicable. 
For \emph{NetConfEval}, we employed its 'low-level configuration generation task', which most closely aligns with \emph{ReGeNet's} intent implementation. 
To ensure a fair comparison, all frameworks used GPT-4.1-mini as the underlying \ac{LLM} and were evaluated on our 10 benchmark scenario dataset (reffered to as \emph{Original} in \Cref{sec:data}) using our automatic evaluation framework (\Cref{sec:llm_eval}). 
Each scenario was executed five times, resulting in 150 solutions across 30 test cases.

Minor compatibility adaptations were performed to align framework-specific inputs and outputs with our benchmark and ensure fair comparison, while preserving the original intent implementation logic.
These adaptations include: \ac{LLM} instructions regarding our JSON topology representation format, output format standardization, and the removal of framework-specific auxiliary information that was not applicable to our Cisco-based scenarios.

\subsection{Solution Quality Criteria} \label{sec:quality}

\subsubsection{Topology Image Understanding Score (TIUS)}\label{sec:topology-understanding-score}

We briefly report the \ac{TIUS}, introduced in our prior work~\cite{ifland2025genet}, to facilitate comparison with earlier results. The \ac{TIUS} measures the accuracy of the \emph{Topology Understanding} module by comparing the generated topology JSON with a manually constructed ground-truth representation. It evaluates the correctness of nodes, node labels, node icons, links, and link-node connections using a weighted sum:
\begin{equation}
    TIUS=0.3\cdot N+0.2\cdot NL+0.05\cdot NI+0.35\cdot L+0.1\cdot LNL
\end{equation}
\revised{where $N$ is the node accuracy score, $NL$ the node-label accuracy score, $NI$ the node-icon accuracy score, $L$ the link accuracy score, and $LNL$ the link-node-label accuracy score, each computed as the fraction of correctly identified elements relative to the ground truth. The weights (0.3, 0.2, 0.05, 0.35, 0.1) reflect the relative importance of each element type and were set heuristically in~\cite{ifland2025genet}.}

However, because these weights are subjective and may not generalize across contexts, we use the DL\_Score introduced in the next subsection as the primary evaluation metric throughout the remainder of the paper.

\subsubsection{Damerau-Levenshtein JSON Edit Distance Score} \label{sec:dl_score}
Inspired by Angi et al.~\cite{angi2025llnet}, who used the Damerau-Levenshtein JSON edit distance similarity score~\cite{boytsov2011indexing} to measure the accuracy of intent translation in \ac{SDN} environments, we adapted this metric to objectively assess the accuracy with which the topology images were interpreted by the \emph{Topology Understanding} module. 
This score quantifies the minimal sequence of operations (deletion, insertion, substitution, and transposition) necessary to transform one string into the other and has been widely used in the literature~\cite{zhao2019string,anderson2024damerau}.
Formally, this metric is denoted recursively as:
\begin{equation}
\small
d_{a,b}(i,j) = \min
\begin{cases}
0, & i = j = 0 \\[1pt]
d_{a,b}(i-1,j) + 1, & i > 0 \\[1pt]
d_{a,b}(i,j-1) + 1, & j > 0 \\[1pt]
d_{a,b}(i-1,j-1) + \scriptstyle 1_{(a_i \neq b_j)}, & i,j > 0 \\[1pt]
d_{a,b}(i-2,j-2) + \scriptstyle 1_{(a_i \neq b_j)}, & 
\begin{aligned}[t]
i,j > 1, \\ a_i = b_{j-1}, \\
a_{i-1} = b_j
\end{aligned}
\end{cases}
\end{equation}
where $a$ and $b$ are the compared strings, and $i$ and $j$ are the lengths of their corresponding substrings.

Specifically, to align this score with the exact match setting, we adapted it to enable it to measure similarity instead of measuring distance using the following formula:
\begin{equation}
s_{a,b}(i,j) = 1 -  d_{a,b}(i,j)
\end{equation}
The score takes values in the range $[0,1]$, where 1 denotes identical JSON strings and 0 denotes completely different JSON strings. Following the approach of Angi et al., we employed LangChain’s tool~\cite{JsonEditDistance} and applied normalization (i.e., customized sorting) to the elements of the JSON structures (lists, dictionaries, and tuples). This ensures proper alignment between the topology JSON representation generated and its corresponding ground-truth JSON representation. In the remainder of the paper we refer to this score as the DL\_Score.

\subsubsection{\Acf{IIS}} \label{sec:fst-scrn}
Each scenario comprises an intent and its associated network specifications. 
To evaluate the effectiveness of the intent's implementation, we developed scenario-specific scoring keys. \revised{These keys were introduced in our previous work~\cite{ifland2025genet} and are based on scenario solutions from GNS3 Vault for configuration scenarios and insights from IT expert interviews for topology scenarios. The keys were refined during the evaluation of GeNet (the predecessor system) to account for common implementation mistakes and valid alternative solutions; whenever a new valid approach or recurring error was identified, the keys were updated and all GeNet results were re-evaluated for consistency. Once ReGeNet development began, the scoring keys were held fixed and were not further modified. This means that prompt refinement, judge-prompt selection, and scoring key development all occurred on the same ten scenarios that produce every reported result; we acknowledge this as a limitation, as it precludes an in-sample/out-of-sample comparison, and note that out-of-sample evaluation on independently constructed scenarios remains an important direction for future work.}

The scoring scheme was designed independently by a team member not involved in the development of either \emph{GeNet} or \emph{ReGeNet.} 
This separation between evaluator design and system implementation was intentionally maintained to ensure that the scoring criteria reflect objective assessment standards rather than system-specific behaviors.

The scoring system primarily assesses \emph{ReGeNet}'s overall performance, and it fundamentally evaluates semantic intent implementation correctness. 
Each evaluation key corresponds to an operational requirement that the generated configuration and topology must satisfy to correctly implement the user's intended network behavior.
For example, consider the following scenario-specific scoring keys:

\begin{example}[Adding Local PCs]
\label{ex:scoring}
For the scenario in Example \ref{ex:scenario}, the suggested scoring keys are: 
\begin{enumerate}
    \item Adding a switch to the topology: 20 points. 
    \item Adding 3 PCs: 10 points per PC.
    \item Correct subnet assignment: 3 points.
    \item Configuring the new switch: 15 points.
\end{enumerate}
\revised{The raw points for this example sum to 68 (20 + 30 + 3 + 15). The \ac{IIS} is normalized to a 0--100 scale by dividing the achieved points by the scenario's total possible points and multiplying by 100; each scenario has a different total, so the denominator varies across scenarios.}
\end{example}
Points are fully or partially deducted for non-compliance, depending on the severity of the deviation.

A detailed breakdown of the scoring keys and point allocations can be found in GeNet's Git repository\footnote{https://github.com/networkcopilot/ICDCS25-GeNet/tree/main/Dataset}, inside each scenario type directory.

\subsection{Scenario Complexity Survey}
\label{sec:survey}
To measure the scenario complexity, we designed a survey which contains seven multiple-choice questions per scenario (Table~\ref{tab:questionnaire}). The survey was distributed to three IT experts with diverse backgrounds and experience levels. Their responses were used to categorize the scenarios and assess their complexity (Section~\ref{sec:survey_res}).

\begin{table}[h!]
\centering
\begin{tabular}{|c|p{7.7cm}|}
\hline
\rowcolor{gray!20}
\textbf{\#} & \textbf{Questions} \\ \hline
1 & How would you rate the overall complexity of this scenario compared to a typical daily task? \\ \hline  
2 & What are the most challenging aspects of this scenario? \\ \hline  
3 & How many working hours would it take to complete the task at hand? \\ \hline  
4 & What level of domain knowledge is needed to solve this task? \\ \hline  
5 & Which specific areas of domain knowledge are important for solving this task? \\ \hline  
6 & How likely are mistakes to occur when performing this task? \\ \hline  
7 & How many test cases are required to validate this scenario? \\ \hline
\end{tabular}
\caption{\centering Survey questions assessing scenario complexity.}
\label{tab:questionnaire}
\end{table}

\subsection{Results}
\label{sec:results}

\subsubsection{Initial Experiment}
\label{sec:initial_results}
As seen in Figure \ref{fig:scores_vs_temperatures}, the temperature levels had minimal impact on the \ac{TIUS} and \ac{IIS}. This suggests that the temperature parameter (0 to 1) does not significantly affect \ac{LLM}-based networking co-pilot performance. For the full experiment, we therefore used a temperature of 1, which is the default setting for the Assistants API (Section~\ref{sec:exp-setup}).  

\begin{figure}[h]
    \centering
    \includegraphics[width=0.90\linewidth]{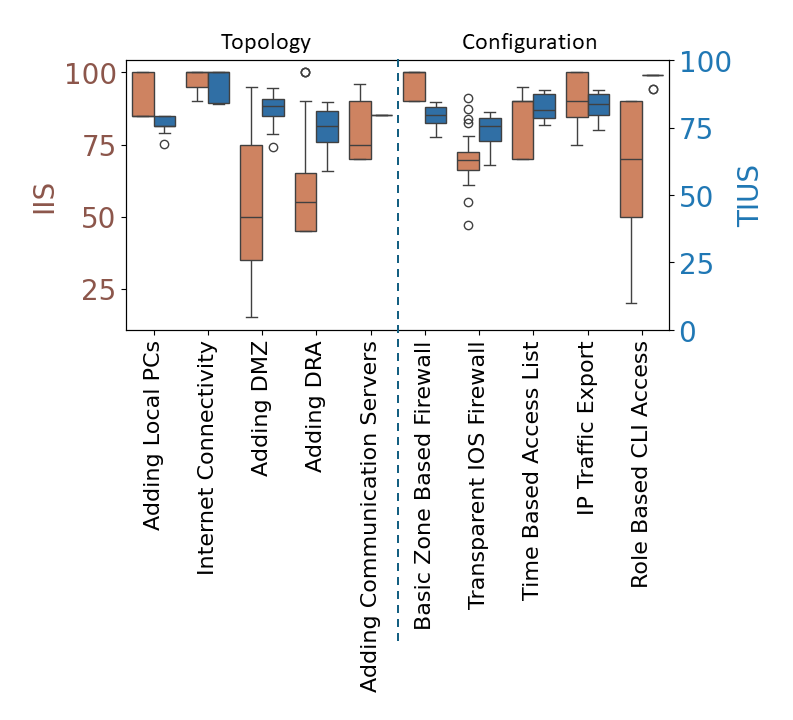}
    \caption{Initial experiment: \ac{TIUS} and \ac{IIS} by scenario.}   
    \label{fig: scores vs scenarios}
\end{figure}

However, as can be seen in Figure \ref{fig: scores vs scenarios}, the \ac{TIUS} and \ac{IIS} values vary by scenario.
In this experiment, one of the topology images lacked labels on the edges indicating the interfaces, which resulted in relatively lower values for the \ac{TIUS} and \ac{IIS}. 
We hypothesize that providing interface information allows the \ac{LLM} to better understand component relationships and implement intents more effectively. This motivated the design of our Vision-Extended dataset for the full experiment (Sections~\ref{sec:data} and~\ref{sec:full_exp}).  
Manual evaluation of the initial \textit{GeNet} solutions showed that of the 250 test cases, \textit{GeNet} correctly configured 79\% of the expected components, reconfigured 92\% of the components requiring updates, and successfully configured 72\% of the newly added components. This highlights the greater challenge of configuring components from scratch.

\subsubsection{Automatic \ac{LLM}-Based Evaluation} \label{sec:llm_eval_res}
\begin{figure}[t]
    \centering
    \includegraphics[width=0.50\linewidth]{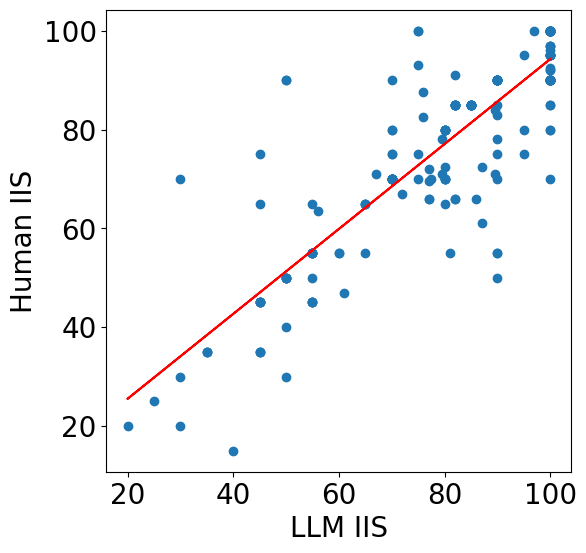}
    \caption{Correlation between human and \ac{LLM} \ac{IIS} evaluations.}
    \label{fig: human vs LLM}
\end{figure}

To align the \ac{LLM}s' scoring with human grading, we experimented with multiple prompts. Initially, we instructed the \ac{LLM} to produce a final score per \textit{GeNet} solution, without an explanation. This approach yielded low correlation (Spearman $<0.5$), most likely due to the complexity of the task.  
To improve correlation, we decomposed the evaluation task into individual scoring criteria.
Finally, we implemented the chain-of-thought prompt strategy, which requires a brief explanation for each IIS scoring key. This significantly increased the Spearman correlation to $0.847$ ($p=4.11 \times 10^{-67}$, Figure \ref{fig: human vs LLM}), demonstrating strong alignment with human grading.

Our analysis identified three sources of discrepancies between human and \ac{LLM} evaluations: hallucinations by the \ac{LLM}, human grading errors, and pipeline errors from the \textit{Topology Understanding} module. The correlation could be improved through further refinement.

\subsubsection{Vision Benchmark (RQ~\ref{rq:images} and RQ~\ref{rq:efficiency})} \label{sec:vision_bench_results}

\begin{figure}[h]
    \centering
    \includegraphics[width=1\linewidth]{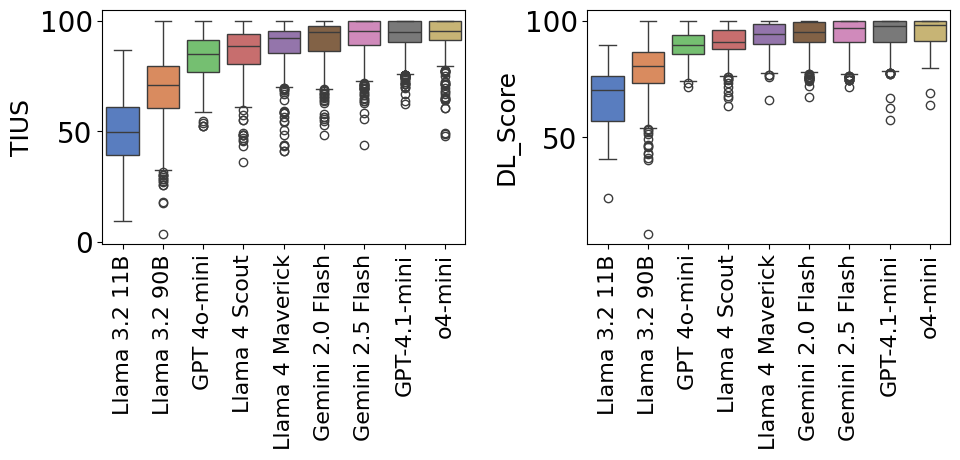}
    \caption{Vision Bench: sorted \ac{TIUS} and DL\_Score by model.}   
    \label{fig: vision_bench_tius_dl}
\end{figure}
\begin{figure}[h]
    \centering
    \includegraphics[width=1.0\linewidth]{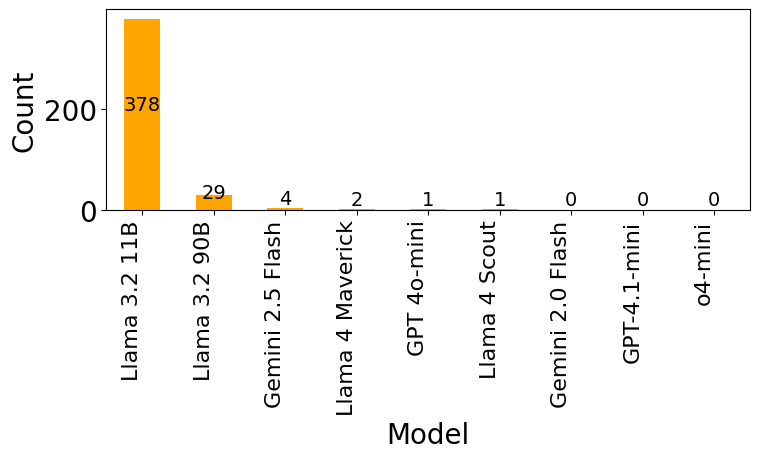}
    \caption{Comparison of the models' failure to produce valid topology JSONs.}   
    \label{fig: vision_bench_failures}
\end{figure}

\paragraph{Performance Across Models}
As seen in \Cref{fig: vision_bench_tius_dl}, \textit{o4-mini} and \textit{GPT-4.1-mini} outperform the other models, while \textit{Llama 3.2 11B} performs the worst. 
All models differ significantly in terms of their mean \ac{TIUS} value, and only \textit{o4-mini} and \textit{GPT-4.1-mini} are indistinguishable based on their mean DL\_Score. 
Moreover, as seen in \Cref{fig: vision_bench_failures}, \textit{Llama 3.2 11B} fails in 84\% of the cases due to invalid JSON outputs, while the other models have negligible failure rates. 
Finally, there is a strong correlation between the DL\_Score and \ac{TIUS} ($\rho=0.829$, $p=1.224 \times 10^{-195}$). Given this, along with the \ac{TIUS}'s previously mentioned drawback (\Cref{sec:topology-understanding-score}), we focus on the DL\_Score in the remainder of our analysis.

To ensure the cost-effectiveness of our experiments, and based on the above results as well as related work highlighting the superior performance of GPT-4 series models on networking tasks, most of our subsequent experiments and analyses focus on the \textit{o4-mini} and \textit{GPT-4.1-mini} models; these models are comparable, since OpenAI reports that they share the same knowledge cutoff.

\begin{figure}[t]
 \centering
\begin{subfigure}{.49\linewidth}
    \centering
    \includegraphics[width=1\linewidth]{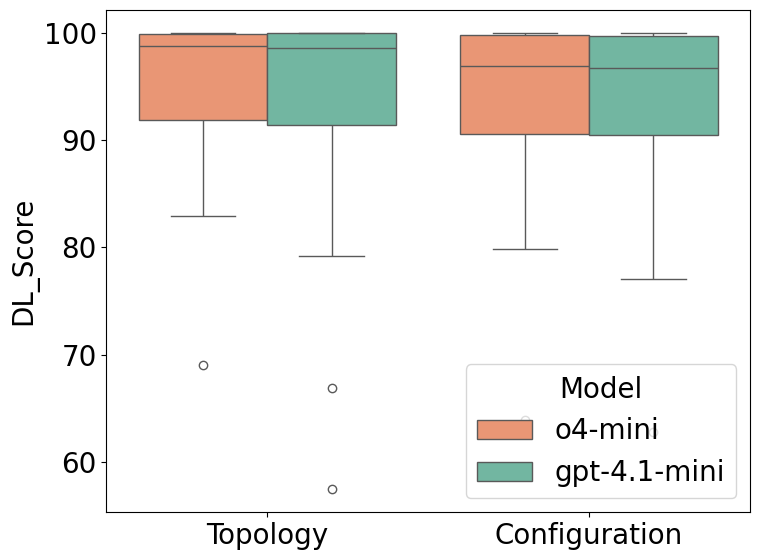}  
    \caption{}
    \label{fig:TIUS vs type - full}
\end{subfigure}
\begin{subfigure}{.49\linewidth}
    \centering
    \includegraphics[width=1\linewidth]{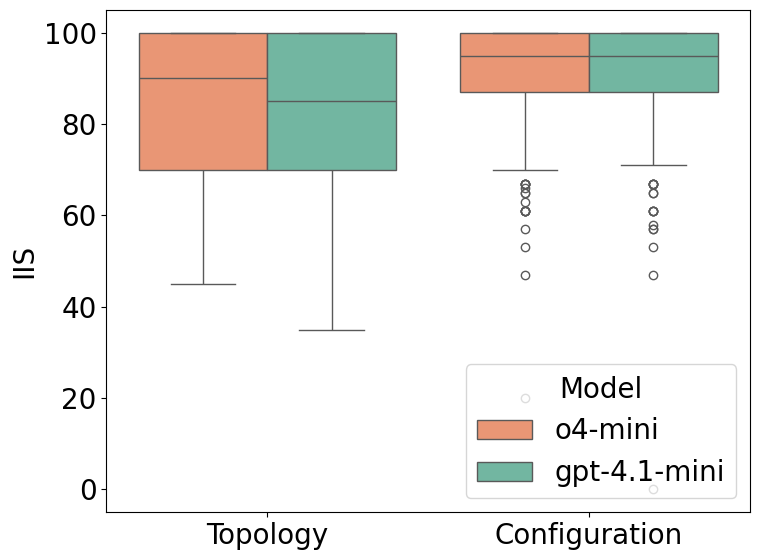}  
    \caption{}
    \label{fig:IIS vs type}
\end{subfigure}
\caption{Scenario type effect on the (a) DL\_Score and (b) \ac{IIS}.}
\end{figure}

Both models achieve mean DL\_Score \revised{$>0.94$} (on the 0--1 scale), even on the least comprehensible topology image variants (\emph{Messy Layout-Paper Sketch}), confirming the potential of utilizing the visual modality in \ac{LLM}-based co-pilots~\cite{wang2024netconfeval}. 
As seen in \Cref{fig:TIUS vs type - full} and backed by the results of statistical tests, the scenario type impacts the DL\_Score. Further statistical analysis reveals that the scenarios accounted for the largest share of variance in the DL\_Score, which can be explained by the varying difficulties of the underlying topologies (e.g., number of nodes and links). For example, the topology in the "Adding DRA" scenario is substantially more complex than that of the "Time-Based Access List" scenario, as can be seen in our Git repository.\footnote{https://github.com/networkcopilot/Scenarios}

\begin{figure}[ht]
    \centering
    \includegraphics[width=1.0\linewidth]{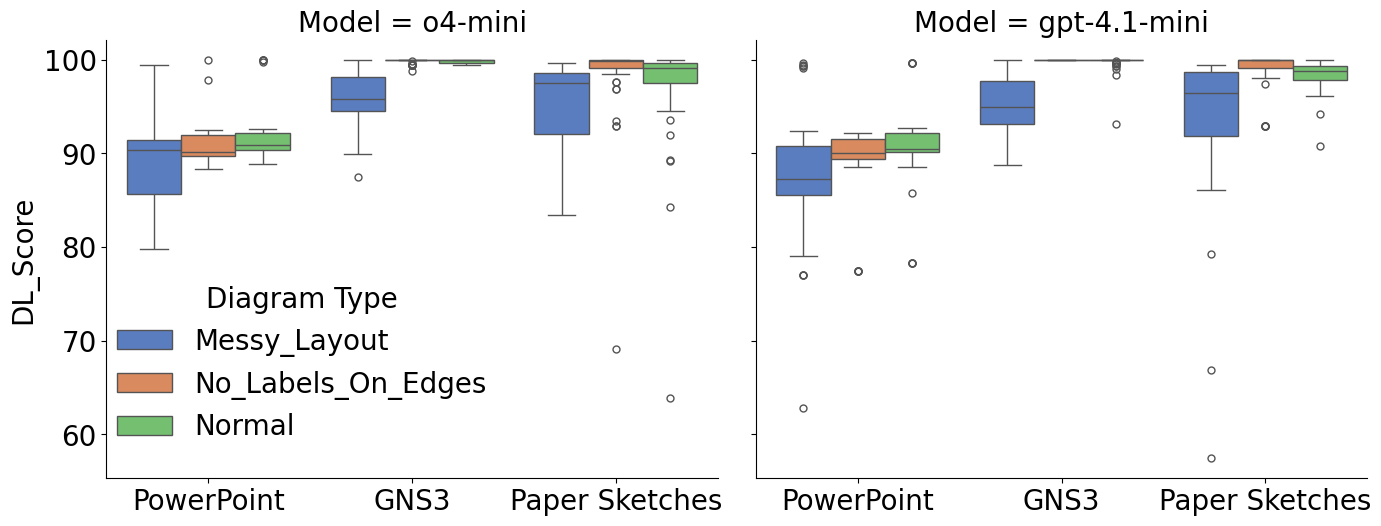}
    \caption{DL\_Score by format and image type per model.}   
    \label{fig: vision_boxplots}
\end{figure}

\paragraph{Visualization Formats and Image Types}
Visualization format and image type also affected DL\_Score. As \Cref{fig: vision_boxplots} shows, GNS3 images scored significantly higher than PowerPoint or paper sketches. \emph{Paper sketches} expectedly scored lowest, but PowerPoint diagrams performed worse still: lacking device icons, models often inferred them from labels rather than leaving the required empty string—an unintended yet useful behavior, showing \acp{LLM}' ability to infer missing context.
As expected, \emph{Messy Layout} images scored considerably lower than the comparable \emph{Normal} and \emph{No Labels on Edges} images, reflecting their suboptimal visualization characteristics; still, some \emph{Messy Layout} images achieved a perfect DL\_Score.
\begin{figure}[h]
    \centering
    \includegraphics[width=0.5\linewidth]{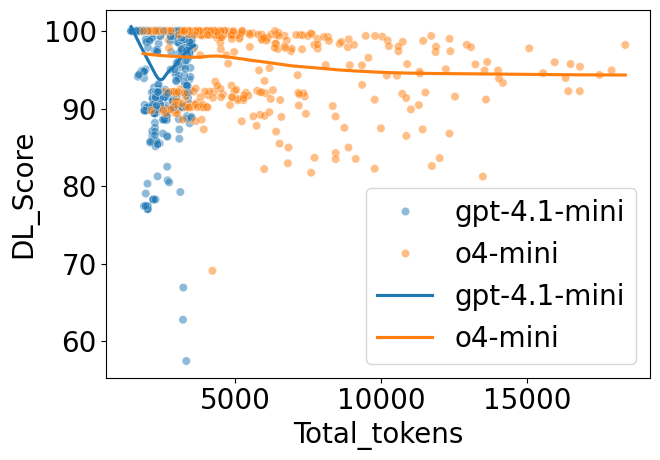}\caption{DL\_Score-token usage trade-off by model.}   
    \label{fig: dl_token_tradeoff}
\end{figure}

\paragraph{Quality-Token Tradeoff}
Since both models achieve statistically similar DL\_Score values, we compare their token usage, directly comparable as both share the same tokenizer. As \Cref{fig: dl_token_tradeoff} shows, \textit{o4-mini}'s DL\_Score stays relatively stable across token counts, while \textit{GPT-4.1-mini} is less stable, dropping around 2,500 tokens; however, \textit{GPT-4.1-mini} is markedly more token-efficient, consistent with \textit{o4-mini}'s reasoning-driven chain-of-thought generating extra tokens. The full nine-model trade-off appears in \Cref{fig: dl_token_tradeoff_facets}.

\begin{figure}[t]
 \centering
\begin{subfigure}{1\linewidth}
    \centering
    \includegraphics[width=\linewidth]{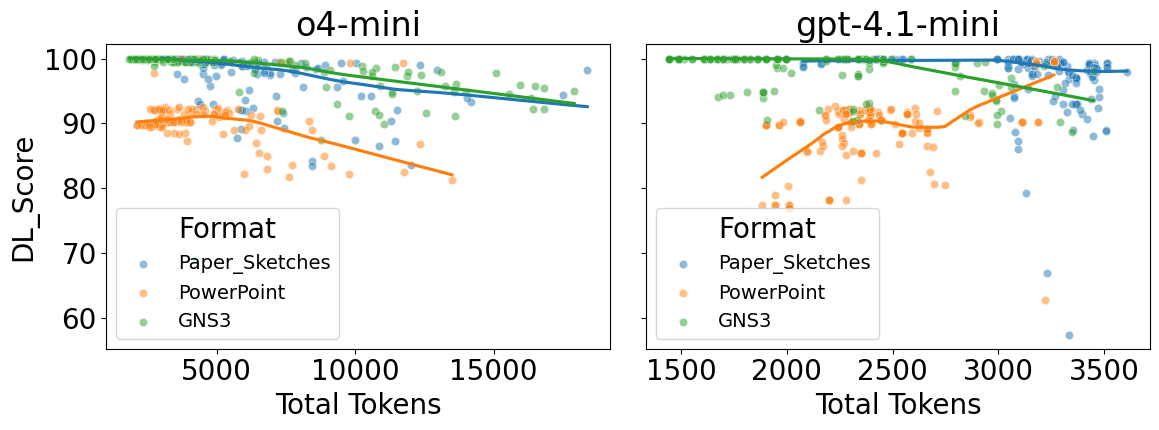}  
    \caption{}
    \label{fig:dl_vs_tokens_platforms}
\end{subfigure}

\begin{subfigure}{1\linewidth}
    \centering
    \includegraphics[width=\linewidth]{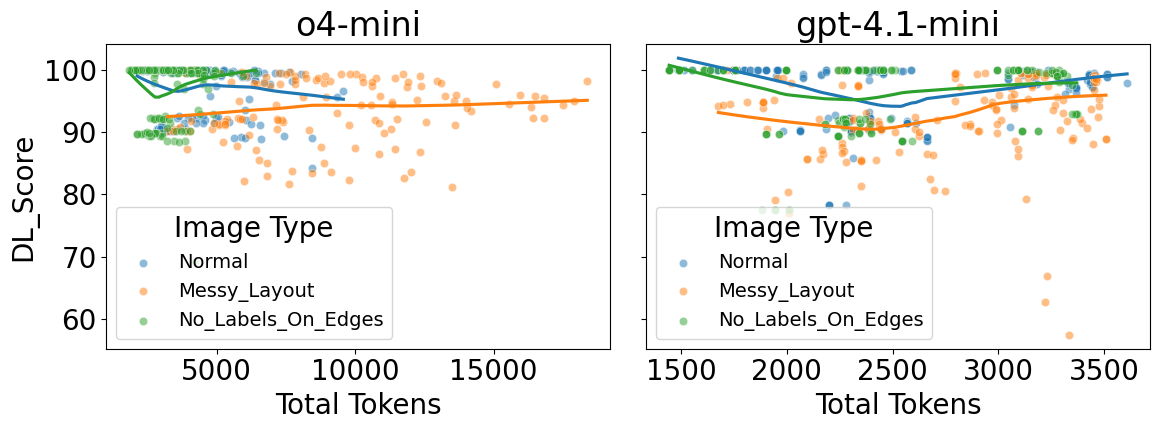}  
    \caption{}
    \label{fig:dl_vs_tokens_diagram_types}
\end{subfigure}

\caption{DL\_Score-token usage trade-off by (a) format and (b) diagram type (b).}
\end{figure}

By format (\Cref{fig:dl_vs_tokens_platforms}), \textit{o4-mini} uses far more tokens without consistently higher quality; per-format analysis shows its DL\_Score declining with token use, especially on PowerPoint topologies, while \textit{GPT-4.1-mini} stays stable or improves with fewer tokens overall. By diagram type (\Cref{fig:dl_vs_tokens_diagram_types}), both models keep a stable DL\_Score, though \textit{o4-mini} again generates far longer outputs, with token usage rising sharply for \emph{Messy Layout} images relative to \emph{Normal} and \emph{No Labels on Edges} types, suggesting less structured inputs prompt longer reasoning. As before, \textit{Messy Layouts} remain hardest to interpret, showing the largest score spread for both models. Overall, \textit{GPT-4.1-mini} matches or exceeds quality with far better token efficiency.

We repeated this analysis using cost instead of tokens; but since cost and tokens strongly correlate (providers charge per token) and \textit{o4-mini}'s per-token cost exceeds \textit{GPT-4.1-mini}'s, we focus on tokens for a fair comparison. Cost plots appear in \Cref{fig:dl_vs_cost_platforms,fig:dl_vs_cost_diagram_types}, with the full nine-model analysis in \Cref{fig: dl_vs_cost_tradeoff_facets}.

\begin{figure}[b]
    \centering
    \includegraphics[width=0.7\linewidth]{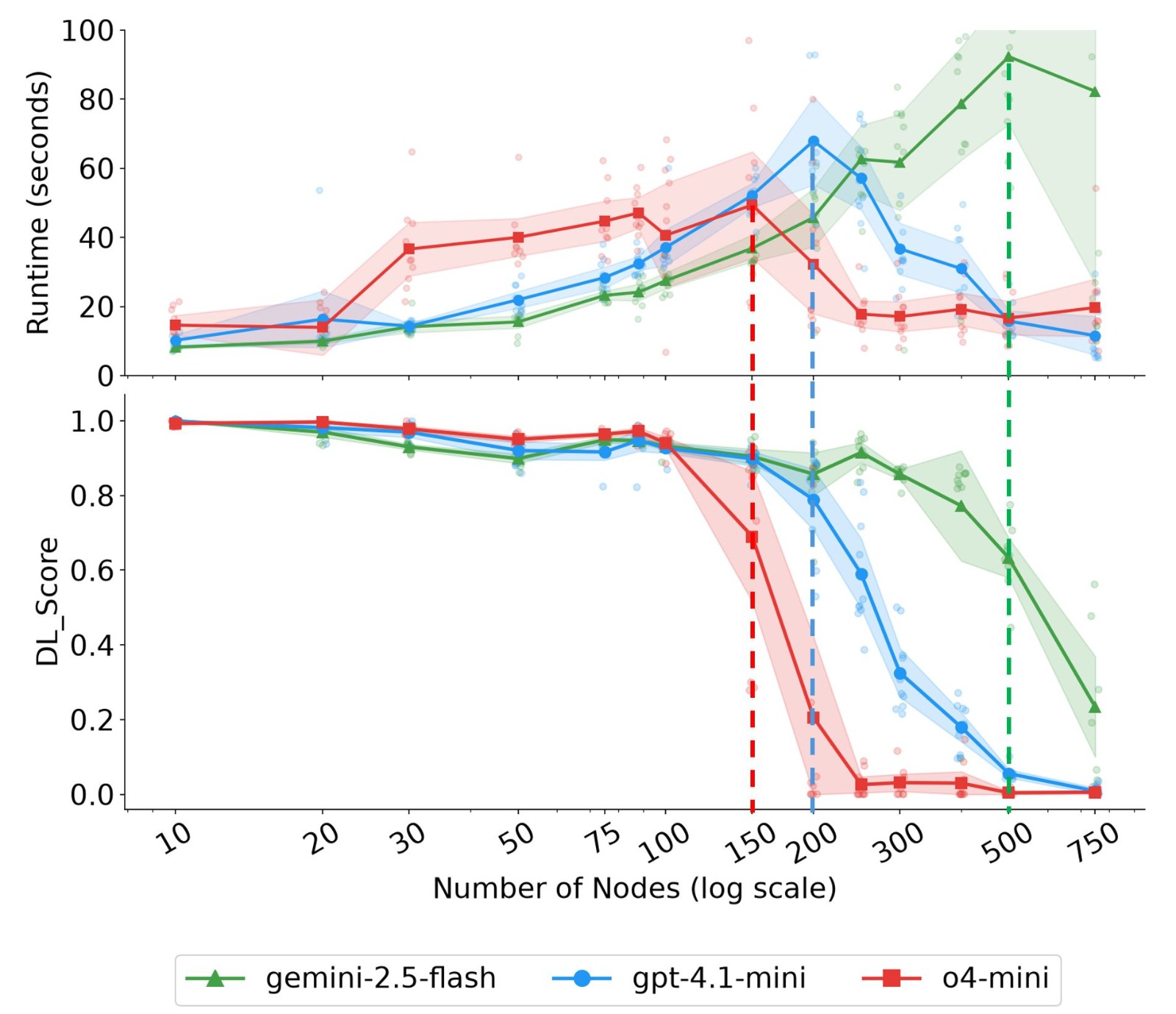}
    \caption{
    DL\_score and runtime VS topology size across models. 
    Vertical lines indicate the empirically identified pivotal topology size for each model.}
    \label{fig:topology@scale}
\end{figure}

\paragraph{Scalability} \label{sec:scalability}
To evaluate the Topology Understanding Module's scalability, \Cref{fig:topology@scale} shows performance and runtime as network size grows from 10 to 750 nodes. All three \acp{LLM} maintain a DL\_Score above 0.9 (on the 0--1 scale) for small and medium topologies, but larger sizes trigger a model-dependent phase transition: runtime rises with topology size up to a model-specific pivotal point—approximately 150, 200, and 500 nodes for o4-mini, GPT-4.1-mini, and Gemini-2.5-Flash, respectively—then declines alongside a sharp drop in DL\_Score, with Gemini-2.5-Flash scaling to substantially larger networks than the others. Invalid or unparsable JSON outputs remained negligible throughout, indicating that scale mainly degrades understanding accuracy rather than output validity. These results suggest current \acp{LLM} handle moderately large enterprise topologies well, though accuracy on very complex visual structures will likely keep improving as context capacity, hierarchical reasoning, segmentation, and \ac{RAG} techniques advance.

\begin{figure}[t]
    \centering
    \includegraphics[width=0.9\linewidth]{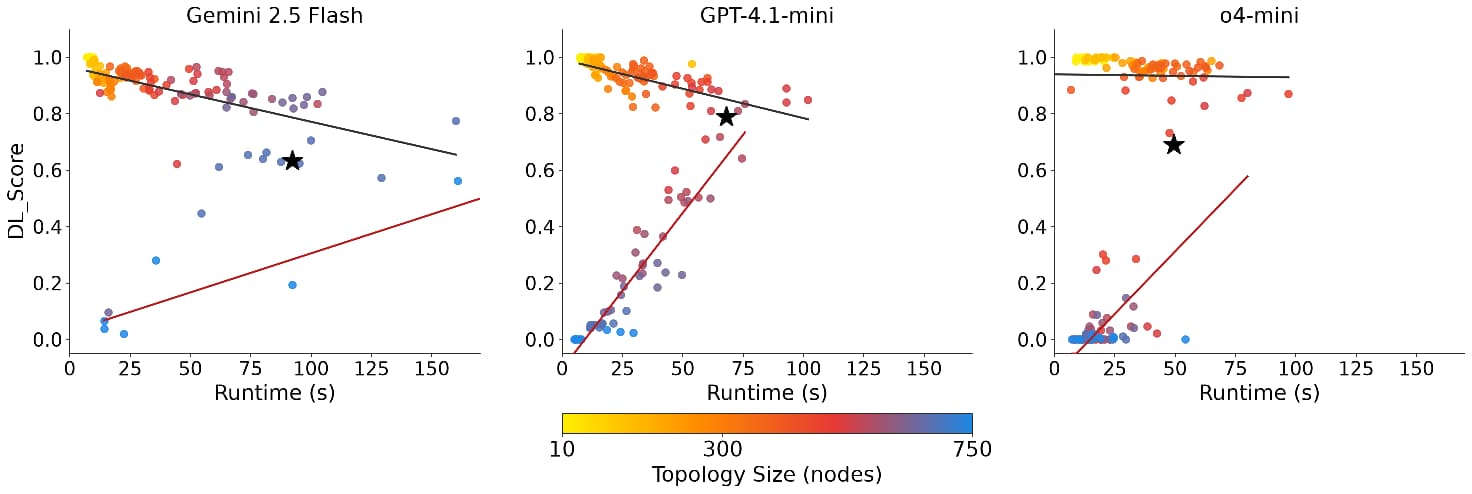}
    \caption{
    DL\_Score VS runtime for the scalability experiment. Two regression lines are used to highlight a scalability behavior transition before and after an empirically identified pivotal point in terms of \#node (marked with a star). 
    Gemini-2.5-Flash: slopes are $-1.9\times10^{-3}$ ($p=6.03\times10^{-9}$) for $\leq500$ nodes  and $2.8\times10^{-3}$ ($p=4.32\times10^{-3}$) for $>500$ nodes.
    GPT-4.1-mini: slopes are $-2.1\times10^{-3}$ ($p=3.54\times10^{-8}$) for $\leq200$ nodes and $1.11\times10^{-2}$ ($p=2.51\times10^{-24}$) for $>200$ nodes.
    o4-mini: slopes are $-1.0\times10^{-4}$ ($p=8.89\times10^{-1}$) for $\leq150$ nodes and $8.9\times10^{-3}$ ($p=1.18\times10^{-11}$) for $>150$ nodes.
     }
    \label{fig:topology@scale_scatter}
\end{figure}

\Cref{fig:topology@scale_scatter} relates runtime to performance: before the pivotal point, runtime grows with topology size as more visual detail is processed, while DL\_score declines only mildly; beyond it, both runtime and DL\_score drop together, suggesting a qualitative shift, possibly toward coarser, more global diagram processing rather than component-by-component reasoning. Practically, this suggests topologies beyond a model's pivotal point should be processed hierarchically—via recursive subnetwork grouping or overlapping image decomposition—to keep each input's visual complexity below the transition point.

\revised{For networks beyond a model's reliable processing limit, a hierarchical decomposition strategy is required. We envision the following pipeline: (1) the full network is partitioned into domain- or layer-based subgraphs (e.g., access, distribution, and core layers, or per-site segments) such that each subgraph falls below the model's pivotal-point node count; (2) each subgraph image is processed independently by the Topology Understanding module, yielding a JSON representation per segment; (3) inter-segment links are identified from a lightweight cross-segment adjacency specification (e.g., a simple list of border interfaces) and used to stitch the per-segment JSON outputs into a globally consistent topology; and (4) the Reflective Intent Implementation module operates on the stitched topology, with the verifier additionally checking cross-segment consistency constraints. This pipeline mirrors the hierarchical or domain-based views that operators already use for large networks, and requires no changes to the core ReGeNet modules. Validating this approach end-to-end on production-scale networks is an important direction for future work (see Section~\ref{sec:conclusion}).}

\subsubsection{Intent Implementation (RQ~\ref{rq: verification_effectiveness} and RQ~\ref{rq: verification_vs_reasoning})} \label{sec:implementation_res}

\paragraph{Scenario Types}
As shown in~\Cref{fig:IIS vs type} and backed by statistical tests, \emph{ReGeNet} performs similarly across the two methods, \emph{GeNet+RV} and \emph{GeNet+V} (i.e., with and without reasoning, respectively), but consistently performs better in configuration scenarios than topology scenarios, regardless of the method used. 
This suggests that the topology update tasks in the dataset are consistently more challenging for \emph{ReGeNet} than the configuration update tasks.
As noted in prior work~\cite{ifland2025genet}, configuration scenarios are generally simpler, as they often involve modifying existing components rather than configuring new ones from scratch. 
In contrast, topology scenarios typically require multiple components to be set up without prior context, demanding a longer and more complex inference process.
Since \acp{LLM} tend to perform better when adapting from existing templates rather than generating solutions entirely from scratch, configuration scenarios may be inherently easier to handle.
In fact, our observations from our initial experiment support this assumption (Section \ref{sec:initial_results}). 
The results of our survey (Section \ref{sec:survey_res}) also support this; on average, experts ranked the configuration scenarios as easier than the topology scenarios.

\begin{figure}[t]
    \centering
    \includegraphics[width=1\linewidth]{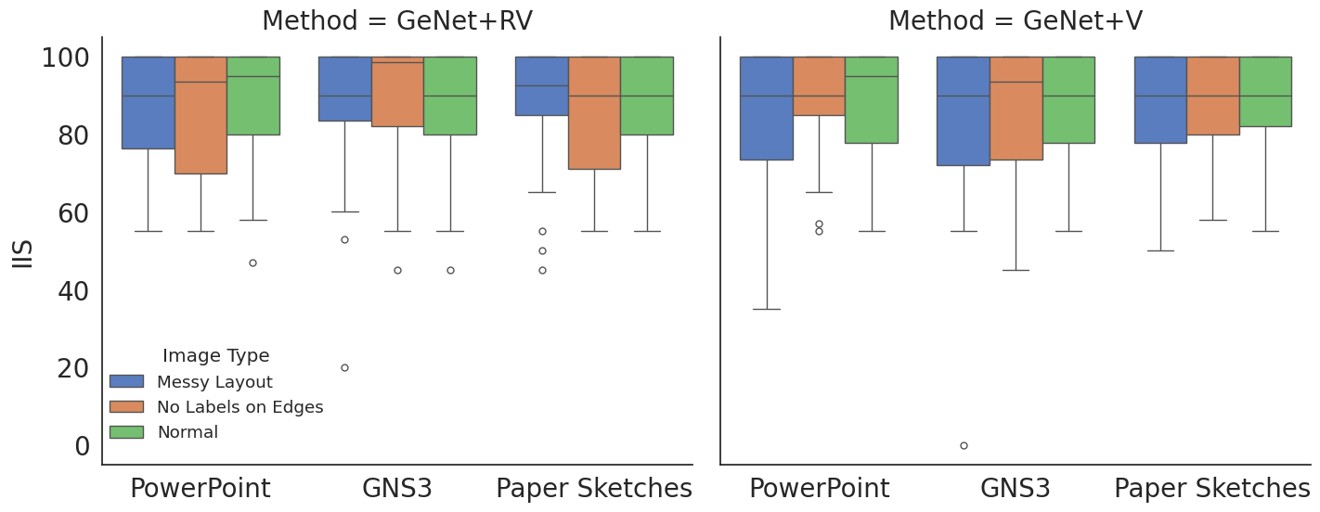}
    \caption{\ac{IIS} by visualization format and image type per method.}   
    \label{fig: IIS_boxplots}
\end{figure}

\paragraph{Visualization Formats and Image Types}
As seen in \Cref{fig: IIS_boxplots} and supported by statistical tests, there are no differences in the \ac{IIS} between the \textit{ReGeNet} methods and within them across different format-image type combinations.
Moreover, among all variables examined, the visualization formats and image types
accounted for the least variance in the \ac{IIS}, whereas the scenario variable 
accounted for the most.
These findings highlight \emph{ReGeNet}'s robustness in handling network specifications of varying quality (e.g., messy layouts) or with missing information (e.g., no labels on edges).
Interestingly, topology images without interface labels on edges seem not to have a significant impact on \emph{ReGeNet}'s intent implementation compared to images containing edge labels. 
This refutes our initial hypothesis regarding edge labels from our initial experiment (Section \ref{sec:initial_results}).

\begin{figure}[t]
\centering
\begin{subfigure}{.48\linewidth}
    \centering
    \includegraphics[width=1\linewidth]{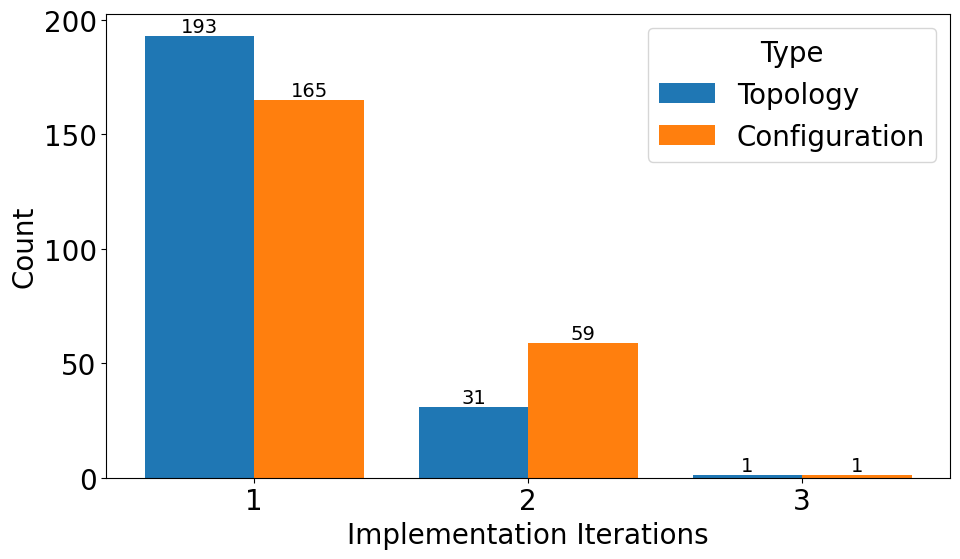}  
    \caption{}
    \label{fig:GeNet+V_iterations}
\end{subfigure}%
\hfill
\begin{subfigure}{.48\linewidth}
    \centering
    \includegraphics[width=1\linewidth]{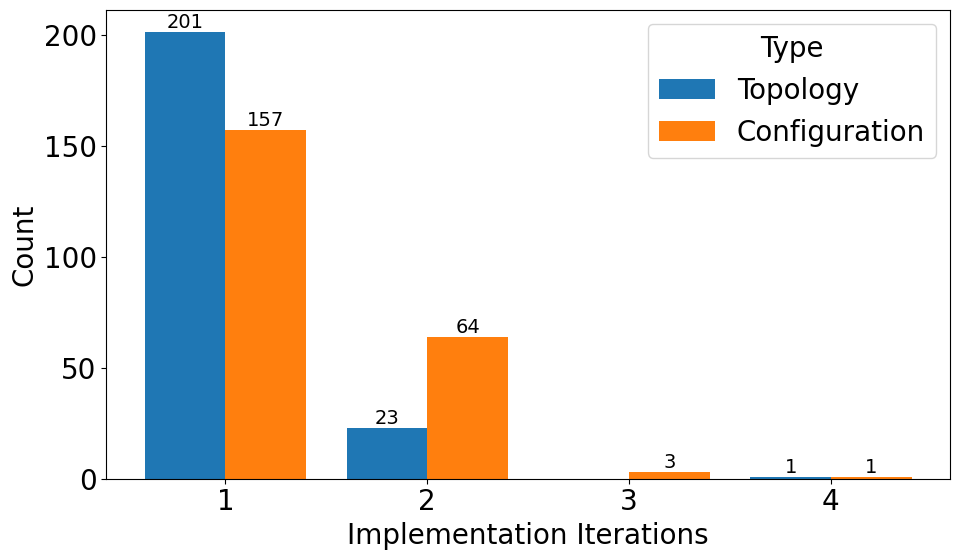}  
    \caption{}
    \label{fig:GeNet+VR_iterations}
\end{subfigure}
\caption{Number of implementation iterations for (a) \emph{GeNet+V} and (b) \emph{GeNet+RV}.}
\label{fig:GeNet_iterations}
\end{figure}

\paragraph{Verification Distribution}
Moreover, since the verification agent in the \emph{Intent Implementation} module determines whether or not to approve a \emph{ReGeNet solution}, the number of implementation iterations varies across cases. 
As shown in~\Cref{fig:GeNet+V_iterations,fig:GeNet+VR_iterations}, in 20\% of the cases, the verification agent did not approve the implementation and provided feedback on the intent implementation agent's solutions. 
Of those, 33\% of the intent implementations processed with \emph{GeNet+V} showed improvement in the \ac{IIS} compared to 27\% for \emph{GeNet+RV}.
Additionally, cases from the topology scenarios exhibited significantly lower improvement rates compared to those from the configuration scenarios, showing 9\% versus 45\% for \emph{GeNet+V} and 4\% versus 35\% for \emph{GeNet+RV}, respectively.
This is consistent with our earlier finding that topology scenarios are inherently more challenging.

\paragraph{Comparison with SOTA Frameworks}
\label{sec:compare_sota}

\revised{We compared \emph{ReGeNet} against two representative LLM-based IBN frameworks, \emph{NetConfEval} and \emph{LLM-NetCFG}, on our 10-scenario benchmark (\Cref{sec:data}), following the setup in \Cref{sec:experimenta_compare}: all frameworks received the same \emph{ground-truth topology JSON}, shared the same underlying LLM and evaluation framework, and were run five times per scenario. This comparison isolates differences in \emph{intent implementation quality} and explicitly excludes the Topology Understanding module; accordingly, the results do not reflect ReGeNet's end-to-end multimodal performance, nor do they capture errors that would propagate from visual topology interpretation in a full pipeline deployment. End-to-end results, where the Topology Understanding module output feeds the implementation stage, are reported in Table~\ref{tab:genet_comparison}.} Table~\ref{tab:framework_comparison} reports the resulting \ac{IIS} scores.
\emph{ReGeNet} achieved the strongest overall performance, with its advantage over the baselines widening further on topology scenarios, consistent with its design for holistic \ac{IBN} tasks that jointly modify topology and configurations, unlike the configuration-focused baselines. As before, topology scenarios scored lower across all frameworks, confirming they remain more challenging than configuration-only tasks.

\begin{table}[t]
\centering
\caption{Intent implementation performance comparison of \emph{ReGeNet} with state-of-the-art frameworks, in terms of the \ac{IIS}, overall and across scenario types.}
    \label{tab:framework_comparison}
    \begin{tabular}{lccc}
    \toprule
    \textbf{Framework} & \textbf{Topology Scen.} & \textbf{Config. Scen.} & \textbf{Overall} \\
    \midrule
    LLM-NetCFG~\cite{lira2024large}            & 74.6          & 88.7          & 81.6          \\
    NetConfEval~\cite{wang2024netconfeval}           & 70.4          & 86.5          & 78.5          \\
    \textbf{ReGeNet}      & \textbf{81.6} & \textbf{93.2} & \textbf{87.4} \\
\bottomrule
\end{tabular}
\end{table}

\paragraph{Comparison with ReGeNet Baselines}
\Cref{tab:genet_comparison} compares \emph{ReGeNet}'s variants against the baselines from \Cref{tab:methods_and_baselines} in terms of \ac{IIS}, reported per scenario given the substantial scenario effect noted earlier. As shown in the 'Average' column, \emph{ReGeNet} significantly outperforms baseline GeNet~\cite{ifland2025genet}, reflecting its newer underlying models, refined system prompts, verification feedback loop, and structured \ac{LLM} outputs for topology JSON generation. Our ablation further shows \emph{GeNet+V} significantly improves over \emph{GeNet+} (i.e., with vs.\ without the verification loop), indicating that for non-reasoning \acp{LLM}, verification is critical to intent implementation quality.

\begin{table*}[h]
\centering
\renewcommand{\arraystretch}{1.2} 

\begin{tabularx}{\textwidth}{|
    l | 
    X | 
    X | 
    X | 
    X | 
    X | 
    X | 
    X | 
    X | 
    X | 
    X | 
    X |
}
\hline
\rowcolor{gray!20}
\textbf{Method} & 
\makecell{\textbf{Adding} \\ \textbf{Comm.} \\ \textbf{Servers}} & 
\makecell{\textbf{Adding} \\ \textbf{DMZ}} & 
\makecell{\textbf{Adding} \\ \textbf{DRA}} & 
\makecell{\textbf{Adding} \\ \textbf{Local} \\\textbf{PCs}} & 
\makecell{\textbf{Basic} \\ \textbf{Zone} \\ \textbf{Based} \\ \textbf{Firewall}} & 
\makecell{\textbf{IP} \\ \textbf{Traffic} \\ \textbf{Export}} & 
\makecell{\textbf{Internet} \\ \textbf{Conn.}} & 
\makecell{\textbf{Role} \\ \textbf{Based} \\ \textbf{CLI} \\ \textbf{Access}} & 
\makecell{\textbf{Time} \\  \textbf{Based} \\ \textbf{ACL}} & 
\makecell{\textbf{Transp.} \\ \textbf{IOS} \\ \textbf{Firewall}} & 
\textbf{Average} \\
\hline
GeNet~\cite{ifland2025genet} & 75.037 & 63.981 & 68.148 & 86.333 & 94.481 & 91.111 & 99.444 & 81.481 & 92.814 & 73.277 & 82.61 \\
\hline
GeNet+ & 81.333* & 73.977* & 74.888n & 91.266* & \underline{98.755}n & 90.444n & \underline{100}n° & 90.222*° & 99.333* & 72.377n & 87.26* \\
\hline
GeNet+V & 81.022 ↓ & 75.111n & \underline{75.555}n & 91.644n & 98.644n & 96.955* & 99.111 ↓ & \underline{90.666}* & 99.444n & \underline{74.311}n & \underline{88.246}* \\
\hline
\revised{GeNet+R} & 81.555n & \underline{75.288}n & 72.666n & \underline{94.266}* & 98.244n & 94.222 ↓ & 99.444n & 87.666 ↓ & 99.555n & 73.311n & 87.622n \\
\hline
\revised{GeNet+RV} & \underline{82.133}n & 70.533 ↓ & 72.333n & 91.8n & 98.155n & \underline{98}* & 98.333 ↓ & 89.333* & \underline{99.822}* & 73.888n & 87.433n \\
\hline
\end{tabularx}

\caption{Comparison of the performance of different \emph{ReGeNet} methods and baselines across scenarios in terms of \ac{IIS}. The highest value in each column is underlined. An asterisk (*) denotes a significant improvement compared to the preceding row, ‘n’ indicates no significant difference, and ‘↓’ marks a significant decrease compared to the preceding row. A degree symbol (°) indicates a significant difference between \emph{GeNet+} and \emph{GeNet+R}.
}
\label{tab:genet_comparison}
\end{table*}

For the reasoning \ac{LLM} ablation, \emph{GeNet+RV} shows no significant improvement over \emph{GeNet+R} (i.e., with vs.\ without verification), and both match \emph{GeNet+V}'s performance—suggesting \textit{o4-mini}'s inherent reasoning already rivals the verification mechanism's benefits, helping minimize errors without it, and that combining reasoning with verification does not trivially yield further quality gains. Since \ac{LLM} reasoning is a recent feature with limited study alongside \ac{LLM}-based verification in \ac{IBN}, this points to an open research direction.

While verification generally improves or maintains intent implementation quality, it has limits: in a few cases, marked~$\downarrow$ in \Cref{tab:genet_comparison} (e.g., \emph{Adding DMZ} under \emph{GeNet+RV}), \revised{it decreased \ac{IIS} relative to unverified implementations, with drops ranging from $-$0.31 to $-$4.76 IIS points depending on the scenario and model.}

\paragraph{Error Analysis and Failure Taxonomy} \label{sec:failure_taxonomy}
To understand recurring failure modes, we manually analyzed 91 samples \revised{(covering both \emph{GeNet+V} (40 samples) and \emph{GeNet+RV} (51 samples))} under a unified taxonomy reported separately per model. The subset comprised (i) all 27 cases where verification significantly degraded \ac{IIS}, (ii) all 10 cases failing verification in the first iteration across all five runs, and (iii) 54 additional cases randomly sampled across IIS quartiles and stratified by scenario, with sampling continued until category saturation. The taxonomy was developed through iterative, emergent coding of recurring error patterns into meaningful categories.

The resulting taxonomy (\Cref{tab:Error_taxonomy}) shows most failures stem from a few recurring patterns, alongside rarer, more specialized categories spanning unsupported assumptions (hallucinations), configuration logic conflicts, and security-related decisions. The dominant mode, \emph{Partial Intent Compliance} (34.0\%), reflects correct understanding of the objective but incomplete implementations that omit required steps. The next most common, \emph{Unwarranted Assumptions} (18.7\%), reflects inferring unsupported topology or device properties, yielding coherent but specification-inconsistent outputs.

Two complementary security behaviors also emerged: \emph{Neglect of Security Best Practices} (omitting recommended mechanisms) and \emph{Prioritization of Security Best Practices over Intent} (overriding explicit requirements)---e.g., generating a hashed instead of an explicitly requested plain-text password, observed only after verifier re-evaluation. \revised{From a controllability standpoint, this constitutes an autonomous agent deviating from an explicit operator instruction, which creates a discrepancy between the operator's mental model and the device's running configuration---an operational risk rather than a safeguard. Production deployments should include approval gates or diff-review steps before applying verified configurations, and operators should be alerted whenever the verifier overrides an explicit intent requirement. Guardrails, rollback mechanisms, and blast-radius controls are necessary complements to any LLM-based verification loop.} Though infrequent, these cases show that \ac{LLM} reasoning can both overlook and proactively enforce security depending on context. Finally, \emph{Internally Contradictory Implementations} captures cases where generated commands negated or overrode one another.

\begin{table*}[t] 
\centering
\caption{Failure taxonomy derived from the manual analysis of 91 samples (40 \emph{ReGeNet+V} and 51 \emph{ReGeNet+RV}). Percentages are reported relative to the analyzed samples of each method and overall. Categories are mutually exclusive; each sample was assigned to the primary failure mechanism that most directly explained the observed implementation error.}
\label{tab:failure_taxonomy}
\renewcommand{\arraystretch}{1.4}

\begin{tabularx}{\textwidth}{p{3.7cm} X X c c c}
\toprule

\textbf{Error Category} &
\textbf{Failure Mechanism} &
\textbf{Typical Manifestations} &
\textbf{ReGeNet+V} &
\textbf{ReGeNet+RV} &
\textbf{Overall} \\

\midrule

\textbf{Partial Intent Compliance} &
Required configuration steps are omitted, resulting in only partial implementation. &
ACL defined but not applied; missing gateway; omitted \texttt{ip inspect}. &
35.0\% &
33.3\% &
34.0\% \\

\textbf{Unwarranted Assumptions} &
Makes unwarranted assumptions about network topology or protocol behavior not implied by the task specification. &
Assumed gateway existence; inferred topology changes; presumed existing connectivity; assumed default protocol behavior. &
15.0\% &
21.6\% &
18.7\% \\

\textbf{Reliance on Implicit Config.} &
Assumes required network behavior without explicit configuration. &
Missing switch configuration; implicit routing assumptions. &
10.0\% &
15.7\% &
13.2\% \\

\textbf{Internally Contradictory Implementations }&
Generates commands that negate or override one another. &
Permit after drop; overwritten routemap; removed applied policy. &
12.5\% &
9.8\% &
11.0\% \\

\textbf{Inappropriate Solution Strategy} &
Selects a technically inappropriate solution that cannot correctly satisfy the intent. &
Misuse of routemaps; incorrect traffic-filtering; inappropriate export strategy. &
7.5\% &
7.8\% &
7.7\% \\

\textbf{Prioritization of Security Best Practices over Intent} &
Overrides explicit task requirements by favoring security best practices. &
Hashed instead of requested plain-text password. &
7.5\% &
5.9\% &
6.6\% \\

\textbf{Neglect of Security Best Practices} &
Omits recommended security configurations despite their relevance to the requested implementation. &
Missing default-deny firewall rule; incomplete DMZ hardening. &
7.5\% &
3.9\% &
5.5\% \\

\textbf{Pipeline Error Propagation} &
Errors introduced during topology understanding propagate to downstream implementation. &
Wrong interface selection; incorrect topology interpretation. &
5.0\% &
2.0\% &
3.3\% \\

\bottomrule

\end{tabularx}
\label{tab:Error_taxonomy}
\end{table*}

To understand when verification degrades implementation quality, we analyzed 32 degradation cases: all 27 cases of significant degradation (e.g., Adding DMZ, see~\cref{tab:genet_comparison}) plus 5 more from the broader failure analysis, sampled until category saturation. \Cref{tab:verification_degradation_taxonomy} presents the resulting mutually exclusive taxonomy of primary verification failure mechanisms, presented in Table~\ref{tab:verification_degradation_taxonomy}.

The most common mechanism, \emph{Misguided Corrective Guidance}, reflects cases where the verifier correctly flagged an issue but suggested an inappropriate fix, lowering quality. The second, \emph{Re-evaluation Inconsistency}, stems from differing evaluator interpretations between the initial and refined implementations rather than genuine verification failure; as noted in \Cref{sec:llm_eval_res}, the evaluator itself can occasionally err or hallucinate, so part of the measured degradation reflects evaluator variability rather than true quality loss. Such evaluator or verifier mistakes could be mitigated via \ac{LLM} validation strategies such as self-refinement~\cite{madaan2023self} and other hallucination-mitigation approaches, potentially paired with human-in-the-loop (HITL) alerts.

\begin{table*}[t]
\centering
\caption{Verification-induced \ac{IIS} degradation taxonomy derived from the manual analysis of 32 cases in which the verification mechanism resulted in a degradation of the \ac{IIS} (\revised{10 GeNet+V and 22 GeNet+RV}).}
\label{tab:verification_degradation_taxonomy}
\renewcommand{\arraystretch}{1.35}

\begin{tabularx}{\textwidth}{
p{4.2cm}
X
>{\centering\arraybackslash}p{1.2cm}
>{\centering\arraybackslash}p{1.2cm}
>{\centering\arraybackslash}p{1.2cm}
}
\toprule

\textbf{Verification Failure Cause} &
\textbf{Failure Mechanism} &
\textbf{ReGeNet+V} &
\textbf{ReGeNet+RV} &
\textbf{Overall} \\

\midrule

\textbf{Misguided Corrective Guidance} &
The verifier correctly identified an implementation issue but proposed an inappropriate corrective action, which led the implementer to produce a degraded solution. &
20.0\% &
36.4\% &
31.2\% \\

\addlinespace

\textbf{Re-evaluation Inconsistency} &
Independent offline evaluations of the initial and refined implementations produced inconsistent judgments due to differing interpretations of \ac{IIS} scoring keys. &
30.0\% &
22.7\% &
25.0\% \\

\addlinespace

\textbf{Security-Driven Behavioral Changes} &
The verifier steered the implementation toward stronger security practices that deviated from the original task requirements. &
30.0\% &
13.6\% &
18.8\% \\

\addlinespace

\textbf{False-Positive Verification Feedback} &
The verifier incorrectly identified a correct implementation as erroneous, prompting unnecessary modifications that degraded the implementation. &
10.0\% &
18.2\% &
15.6\% \\

\addlinespace

\textbf{Context Loss During Re-implementation} &
Previously correct implementation elements were omitted or overwritten during re-implementation despite being unrelated to the feedback. &
10.0\% &
9.1\% &
9.4\% \\

\bottomrule
\end{tabularx}
\end{table*}

\begin{figure}[t]
 \centering
\begin{subfigure}{.30\linewidth}
    \centering
    \includegraphics[width=1\linewidth]{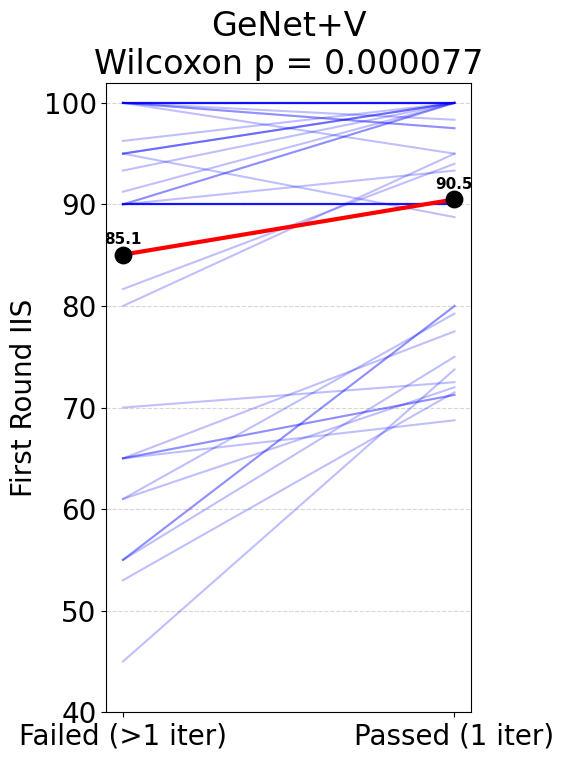}  
    \caption{}
    \label{fig:matched_cohort_4.1}
\end{subfigure}
\begin{subfigure}{.30\linewidth}
    \centering
    \includegraphics[width=1\linewidth]{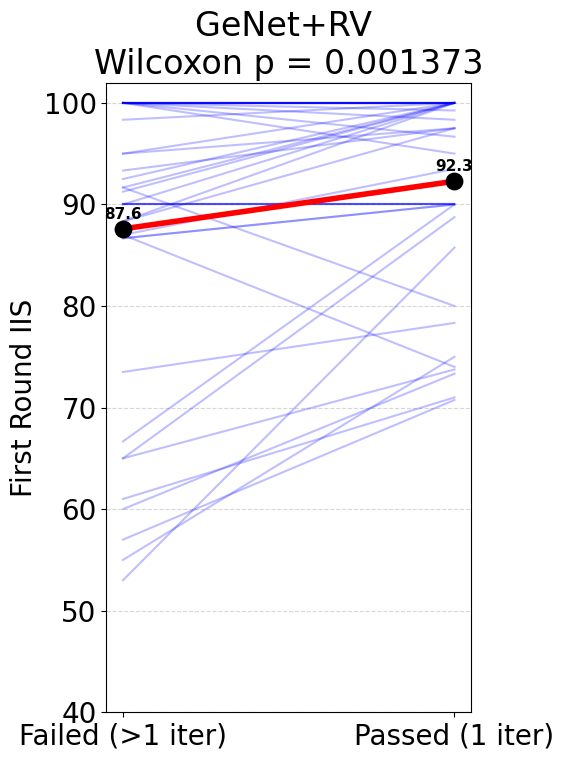}  
    \caption{}
    \label{fig:matched_cohort_o4}
\end{subfigure}
\caption{Paired comparison of first-implementation-round \ac{IIS} per case across iterations that either passed verification in the first implementation round (“Passed (1 iter)”) or required additional correction iterations before passing (“Failed ($>$1 iter)”). 
Each line represents a single matched case, while the red line marks the group means. 
Across both models (a) \emph{GeNet+V} and (b) \emph{GeNet+RV}, iterations passing verification immediately exhibit significantly higher initial \ac{IIS}.}
\label{fig:matched_cohort}
\end{figure}

\paragraph{Deployment Considerations} \label{sec:deployment}
Mission-critical deployments need mechanisms to flag when human oversight is required. We performed a matched-cohort analysis comparing first-round (pre-verification) scores for cases that passed verification immediately versus those needing correction. As \Cref{fig:matched_cohort} shows, immediately-passing solutions scored significantly higher for both models (Wilcoxon signed-rank test, p = 0.0013 for \emph{GeNet+V}, p = 0.000077 for \emph{GeNet+RV}). Although these scores are offline-only and unavailable in production, verification behavior itself may serve as a practical proxy for implementation confidence, flagging non-immediately-passing solutions for optional or mandatory human inspection based on risk tolerance. \revised{We note that this analysis conditions on the verifier's own decision---cases that pass verification are compared against those that do not---which means the result reflects a selection effect rather than a calibrated confidence estimate. Establishing the verifier's precision, recall, and calibration against an externally observable outcome (e.g., parse success or functional reachability) is an important direction for future work.}

While verification-and-correction typically improves final quality over the initial attempt (\Cref{tab:genet_comparison}), repeated failures still signal lower initial confidence and greater corrective effort, potentially warranting escalation to human review in mission-critical settings---e.g., after a single failure in sensitive environments, or only after multiple failures where throughput is prioritized.

\revised{\paragraph{Threat Model and Trust Boundaries}
ReGeNet accepts topology images as input and holds configuration-write authority over device specifications. Topology images are therefore untrusted input: a maliciously crafted diagram---e.g., one containing hidden text labels or adversarially arranged elements---constitutes a viable prompt-injection vector against the Topology Understanding agent. Additionally, device configurations generated by the system (Table~\ref{tab:failure_taxonomy} documents a plaintext-password scenario) are transmitted to third-party API providers, representing a data-exposure risk in sensitive environments. Production deployments should enforce: (i) a trust boundary at the topology image input, with provenance verification where possible; (ii) human-in-the-loop approval before generated configurations are applied to live devices; (iii) rollback and blast-radius controls in the underlying change management system; and (iv) data-handling agreements with API providers governing configuration content. The indirect prompt injection threat is discussed in detail in \cite{greshake2023not}.}

\revised{\paragraph{Limitations and Scope of Evidence}
We note that the benchmark, scoring keys, automatic evaluator (IIS), topology metric (DL\_Score), and the framework being extended all originate in or build directly on our prior work~\cite{ifland2025genet}. No external benchmark or independently constructed evaluation set was used. All reported results should therefore be interpreted in the context of this single evaluation ecosystem, and external validation on independently constructed scenarios and scoring criteria is an important direction for future work.}

\subsubsection{Scenario Complexity vs. Performance (RQ~\ref{rq:questionnaire} and RQ~\ref{rq:efficiency})}
\label{sec:survey_res}

Expert evaluations enabled us to categorize the test scenarios into three levels of perceived complexity: \emph{easy}, \emph{medium}, and \emph{hard} (Table~\ref{tab:scenario_categories}). This categorization allows us to examine how scenario complexity interacts with both the quality and efficiency of \ac{ReGeNet}'s performance.

\paragraph{Topology Understanding and Scenario Complexity}
For the \emph{Topology Understanding} task, expert-rated complexity levels showed no significant effect on the DL\_Score, token usage, or cost. 
This indicates that the \acp{VLLM} used by \ac{ReGeNet} are largely robust to the visual and structural complexity of the topology diagrams.

\begin{figure}[t]
    \centering
    \includegraphics[width=0.7\linewidth]{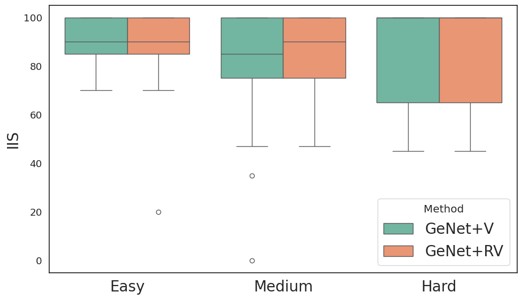}
    \caption{\ac{IIS} by scenario complexity for each method.}   
    \label{fig: IIS vs experts}
\end{figure}

\paragraph{Intent Implementation Quality}
As illustrated in \Cref{fig: IIS vs experts} and backed by statistical analysis, there is not a significant difference between \emph{GeNet+V} and \emph{GeNet+RV} when results were grouped by scenario complexity. 
However, both methods exhibit notably lower \ac{IIS} variance in the \emph{easy} scenarios compared to \emph{medium} and \emph{hard} ones, which show similar variance levels, a predictable yet informative result.
For \emph{GeNet+V}, \revised{we did not find a significant difference in mean \ac{IIS} across complexity levels, suggesting ReGeNet handles complex intents comparably to simpler ones when using a non-reasoning \ac{LLM}; however, given the small number of scenarios per complexity group, this null result should be interpreted with caution rather than as positive evidence of equivalence.} 
In contrast, \emph{GeNet+RV} achieves higher average scores in the easy scenarios relative to medium and hard ones, which exhibit comparable quality.

\paragraph{Efficiency and Token Usage}
Both \emph{GeNet+V} and \emph{GeNet+RV} consume significantly fewer tokens in the \emph{easy} scenarios than in the \emph{medium} and \emph{hard} ones, which have comparable token usage. 
Although \emph{GeNet+RV} generally requires more tokens overall, an analysis limited to completion tokens (i.e., model output rather than internal reasoning) reveals no significant difference between the two models.

Unlike in the topology interpretation task, however, no clear or consistent relationship was observed between input-related factors (e.g., format, diagram type, scenario type, or complexity level) and the quality-token trade-off during intent implementation. 
While scenario characteristics appear to influence this trade-off the most, the effect is not systematic. However, on average, implementation quality tends to slightly decrease as token usage increases.

\begin{figure}[t]
    \centering
    \includegraphics[width=0.6\linewidth]{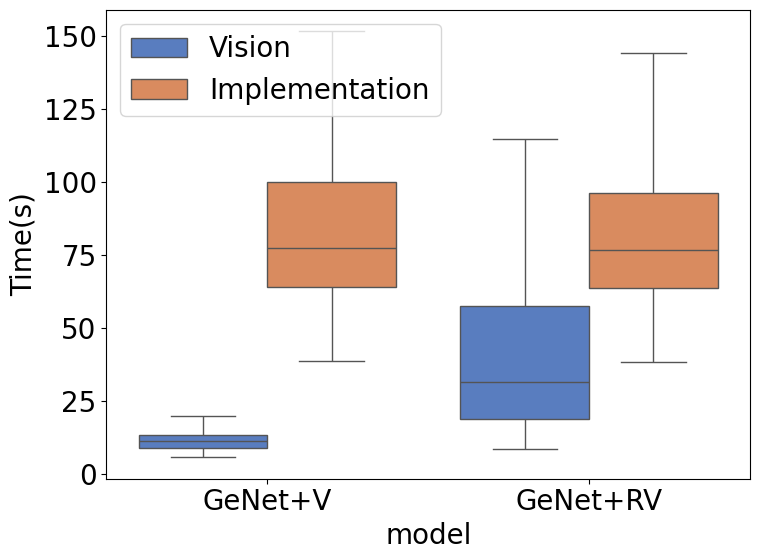}
    \caption{Time distribution of \emph{ReGeNet}'s modules by method. 
    Vision (blue) denotes the \emph{Topology Understanding} module, and Implementation (orange) denotes the \emph{Reflective Intent Implementation} module.}   
    \label{fig: time_boxplots}
\end{figure}

\paragraph{Execution Time and Operational Efficiency} \label{sec:time}
Practical deployment also depends on execution latency. As \Cref{fig: time_boxplots} shows, implementation times are comparable across methods, but topology interpretation ("Vision") is significantly faster and less variable for \emph{GeNet+V} than \emph{GeNet+RV}, likely reflecting \emph{GeNet+RV}'s extra reasoning steps, which do not necessarily improve \emph{Topology Understanding}. Despite this longer inference, overall implementation times remain similar, suggesting non-reasoning \acp{LLM} may offer a better accuracy-efficiency balance for multimodal \ac{IBN} deployment. We compare execution time only between \emph{GeNet+V} and \emph{GeNet+RV}, which share a provider; cross-provider comparisons were not attempted due to differing APIs.
While execution time alone does not capture operational efficiency, it is a prerequisite for practical deployment and suggests potential for reducing effort in repetitive intent implementation.
In summary, this section evaluated \ac{ReGeNet} via experiments addressing our research questions across topology interpretation, intent implementation, and scenario complexity, including a temperature study, automated \ac{LLM} evaluation, and large-scale vision-model benchmarking. The results demonstrate \ac{ReGeNet}'s robustness, the impact of visualization and scenario type, and the effectiveness of reasoning and verification in multimodal intent realization.

\section{Conclusion and Future Work}\label{sec:conclusion}

In this work, we presented \textit{ReGeNet}, a multi-agent framework combining self-verification and reasoning to improve reliability in multimodal \ac{IBN}. Building on \textit{GeNet}, \textit{ReGeNet} adds a reflective verification loop in which implementation and verification agents iteratively refine configurations and topologies until the intent is accurately realized.

Our experiments show that \revised{verification improves LLM-judged intent implementation quality for non-reasoning \acp{LLM} by +0.99 IIS points on average, enabling GeNet+V to match the performance of the reasoning model baseline (GeNet+R). We note, however, that GeNet+ (non-reasoning, no verification) and GeNet+R (reasoning, no verification) are already statistically indistinguishable (average IIS: 87.26 vs.\ 87.62), meaning that the efficient non-reasoning model matches the reasoning model even without verification. Adding verification to a reasoning model (GeNet+RV) yields no significant further improvement ($-$0.19 IIS points), consistent with the self-correction literature~\cite{huang2024large}. The combined effect of verification and reasoning is therefore not additive: the two mechanisms provide partially overlapping benefits, and} the practical implication is that engineers can achieve comparable quality either via a lightweight non-reasoning model with verification or a costlier reasoning model without it. To our knowledge, this is the first systematic study of this interaction in multimodal \ac{IBN}; it implies engineers can achieve comparable quality either via a lightweight non-reasoning model with verification or a costlier reasoning model without it, making reflective verification a cost-effective path to reliable automation rather than something reasoning models render obsolete.

\revised{A natural question raised by the rapid evolution of LLM capabilities is whether a single higher-capacity model could replace the multi-agent loop. We argue that the two are complementary rather than substitutes. First, our results show that GeNet+V---a lightweight non-reasoning model paired with verification---already matches the performance of GeNet+RV, the reasoning model with verification, and does so more cost-efficiently; a practitioner upgrading to a heavier model would still benefit from the explicit correction trace and separation-of-concerns that a two-agent architecture provides. Second, the self-correction literature suggests that same-model verification loops yield diminishing returns as model capability increases~\cite{huang2024large}, but that cross-model verification---where the verifier belongs to a different model family---introduces a qualitatively independent signal that model scale alone cannot replicate. Third, for production IBN deployments, interpretability and auditability of corrections are operationally required independently of raw performance: an engineer must be able to inspect why a configuration was changed, a property that a single-model pipeline does not provide. We therefore expect the architectural benefits of reflective multi-agent verification to persist as underlying models improve, even if the performance gap that justifies the loop for same-family, same-model configurations narrows.}

Reasoning \acp{LLM} remain less efficient in token consumption and execution time, suggesting the optimal integration of reasoning and verification depends on a model's cognitive design and prompting strategy. \textit{ReGeNet}'s multimodal design, jointly interpreting textual and visual network data, shows strong potential for practical, interpretable \ac{IBN} automation.

The framework was not hand-tailored to any vendor syntax and operates directly on raw configurations, topology images, and natural-language intents, making it architecturally applicable beyond Cisco; however, since the underlying \acp{LLM} may be more familiar with Cisco IOS, we do not claim cross-vendor validation here. Similarly, our Cisco-inspired evaluation scenarios, though realistic, derive from GNS3Vault certification exercises rather than full production deployments; while we did assess Topology Understanding scalability on large synthetic enterprise networks, future work should extend end-to-end evaluation to production-scale, multi-vendor, and operationally constrained settings with long-running change management.

\revised{Two further limitations bound the scope of our correctness claims. First, all correctness assessments in this work are based on LLM-judged textual plausibility against scenario-specific scoring keys; no generated configuration was loaded into a network emulator, parsed by a device, or tested with active probes (reachability, ACL hit matrices, or data-plane validation). Functional validation---e.g., replaying configurations through GNS3 or running Batfish checks---would move the central claims from plausible to demonstrated and is a priority for future work. Second, ReGeNet generates CLI-level configurations rather than NETCONF/YANG model-driven changes~\cite{rfc6241,rfc7950}; production intent-based networking deployments typically require model-driven configuration with atomicity, commit confirmation, rollback, and drift detection, none of which are provided by raw CLI text. Integrating with the NETCONF/YANG stack~\cite{rfc9315} is an important direction toward production readiness.}

The observed phase transition in topology understanding further suggests multimodal models shift visual reasoning strategy as network complexity grows, offering a practical estimate of each model's reliable processing limit and motivating hierarchical or tiled topology decomposition beyond that point.

Future work includes: (i) deeper analysis of how reasoning and verification interact to jointly maximize quality and robustness; \revised{(ii) hybrid verification combining LLM intent-alignment reasoning with formal, topology-aware configuration validation (e.g., Batfish or NETCONF \texttt{<validate>}), so that the complementary strengths of both approaches are exploited: the LLM handles ambiguous, partially specified intents while the formal analyzer enforces syntactic and semantic correctness on the generated configurations;} (iii) systematic cross-vendor evaluation (e.g., Juniper, Arista, Huawei) to test the framework's generality beyond Cisco; (iv) applying \ac{LLM}-based verification within the \emph{Topology Understanding} module to reduce cascading errors; (v) controlled user studies quantifying management KPIs such as MTTR reduction and manual effort relative to conventional workflows; \revised{and (vi) end-to-end evaluation of a hierarchical topology decomposition pipeline on production-scale, multi-domain networks (Section~\ref{sec:scalability}), validating the stitching strategy described there and quantifying the error propagation from per-segment understanding to global intent implementation.}

\revised{A deployment consideration not addressed in the current evaluation is the effect of model-level safety guardrails. LLM providers increasingly apply content filters that may refuse to generate or evaluate configurations involving firewall rules, ACL policies, or authentication credentials---content that is routine in network engineering contexts but may superficially resemble offensive security tasks. In our experiments no hard refusals were observed, likely because the scenarios are framed as legitimate IBN tasks; however, this behavior is model- and provider-dependent and should be explicitly tested when deploying ReGeNet with models or providers not evaluated here. Prompt framing (e.g., explicit professional context and scope declarations) and the use of enterprise API tiers with relaxed content policies are practical mitigations.}

Overall, reflective verification is a practical, effective strategy for improving multimodal network copilots, especially alongside efficient non-reasoning \acp{LLM}. More broadly, reliable \ac{IBN} systems require understanding not just the reasoning-verification interplay but also the failure modes and scalability characteristics of multimodal foundation models; to this end, we contribute a failure-mode taxonomy, a verification-induced error taxonomy, and an empirical characterization of topology-understanding scalability, including a model-dependent phase transition with implications for hierarchical topology decomposition, together offering actionable guidelines for reliable, scalable, cost-effective AI-assisted network management.

   

\bibliographystyle{IEEEtran}
\bibliography{bibliography}

\appendix

\subsection{Vision Bench Cost and Full Trade-Offs}
\begin{figure}[t]
    \centering
    \includegraphics[width=1\linewidth]{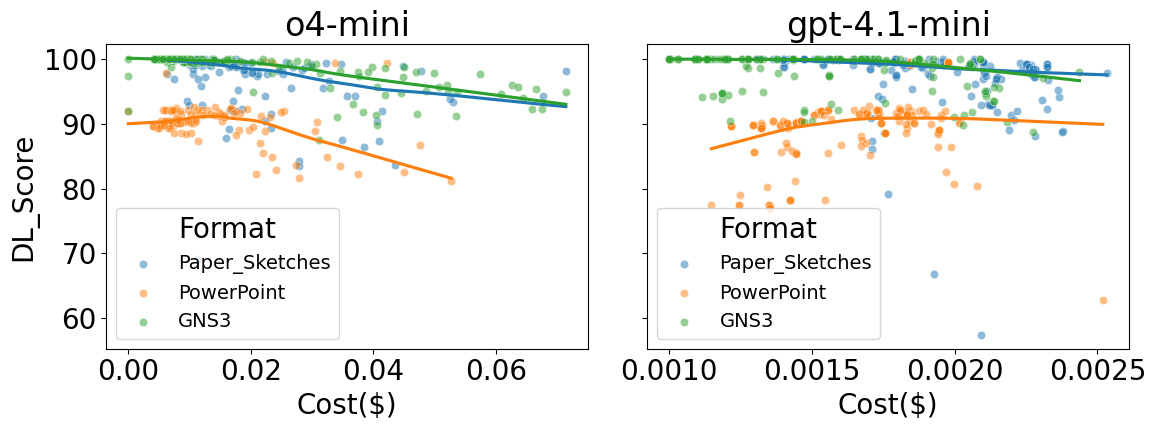}
    \caption{The DL\_Score-cost tradeoff by visualization format.}
    \label{fig:dl_vs_cost_platforms}
\end{figure}

\begin{figure}[t]
    \centering
    \includegraphics[width=1\linewidth]{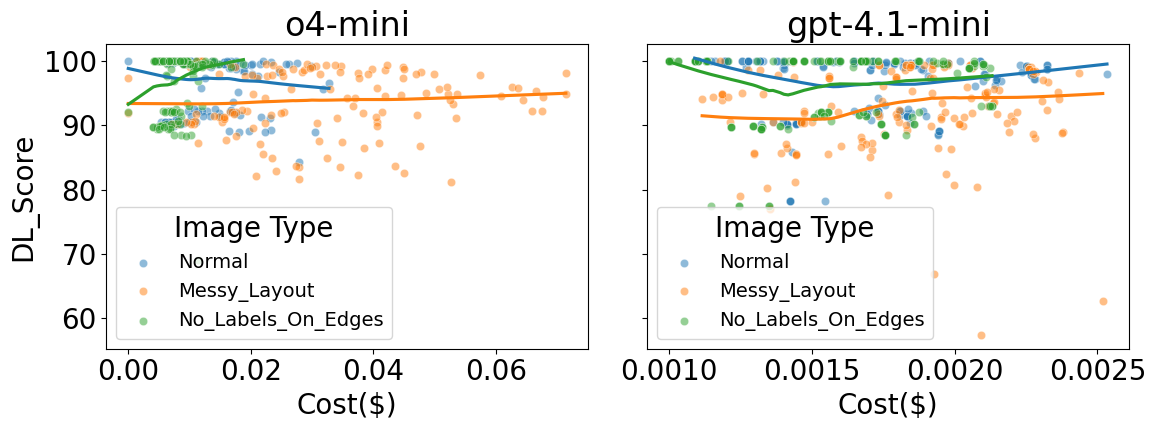}
    \caption{The DL\_Score-cost tradeoff by image type.}
    \label{fig:dl_vs_cost_diagram_types}
\end{figure}

\begin{figure}[h]
    \centering
    \includegraphics[width=0.9\linewidth]{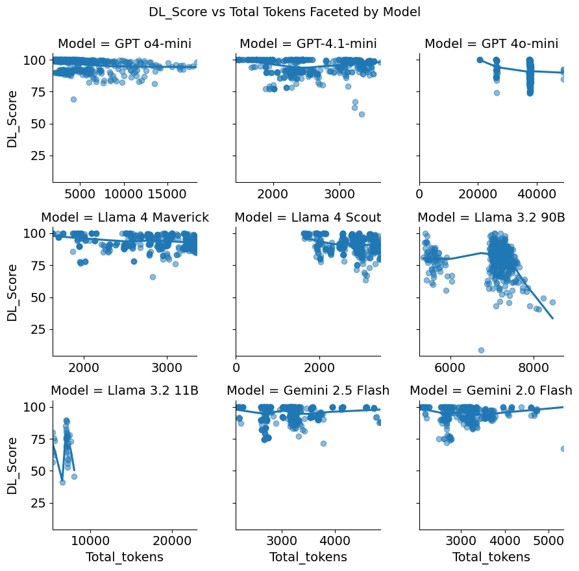}\caption{The DL\_Score-token usage tradeoff for all Vision Bench models.}   
    \label{fig: dl_token_tradeoff_facets}
\end{figure}

\begin{figure}[h]
    \centering
    \includegraphics[width=0.9\linewidth]{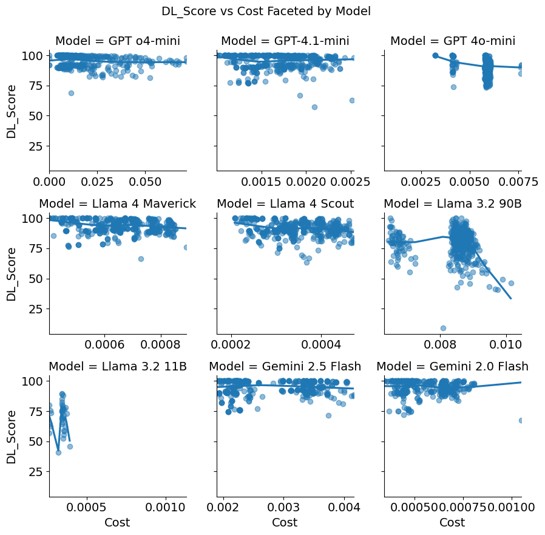}\caption{The DL\_Score-cost tradeoff for all Vision Bench models.}   
    \label{fig: dl_vs_cost_tradeoff_facets}
\end{figure}

\begin{figure}[t]
 \centering
 \begin{subfigure}{.45\linewidth}
    \centering
    \includegraphics[width=\linewidth]{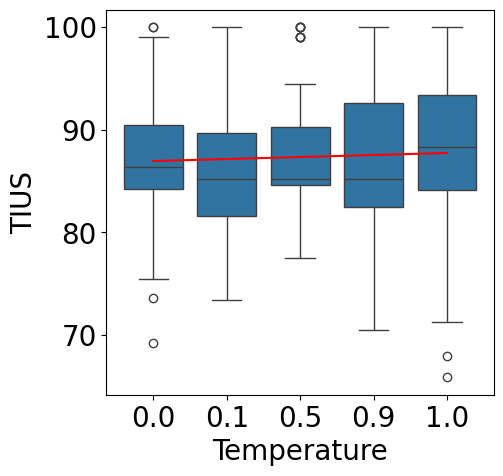}  
    \caption{}
    \label{fig:TIUS_vs_type}
 \end{subfigure}
 \hfill
 \begin{subfigure}{.45\linewidth}
    \centering
    \includegraphics[width=\linewidth]{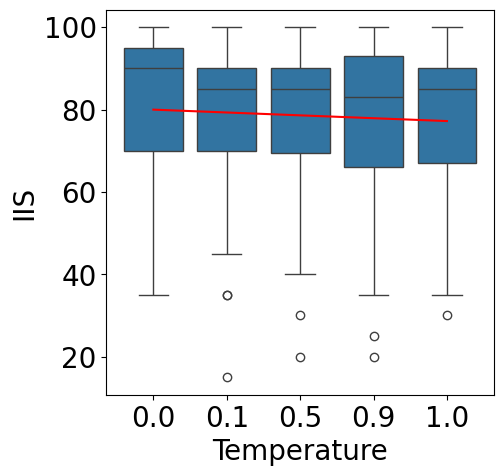}  
    \caption{}
    \label{fig:IIS_vs_temp}
 \end{subfigure}
 \caption{Initial experiment: the effect of temperature on the (a) \ac{TIUS}  and (b) \ac{IIS}.}
 \label{fig:scores_vs_temperatures}
\end{figure}

Figure \ref{fig:scores_vs_temperatures} shows the impact of \ac{LLM} temperature on intent implementation and topology understanding. 
Figures \ref{fig:dl_vs_cost_platforms} and \ref{fig:dl_vs_cost_diagram_types} present the DL\_Score-cost tradeoff for the various visualization formats and image types.
Figures \ref{fig: dl_token_tradeoff_facets} and \ref{fig: dl_vs_cost_tradeoff_facets} present the DL\_Score-token usage and cost tradeoffs for the entire Vision Bench experiment.

\end{document}